# Active microrheology of colloidal suspensions: simulation and microstructural theory

Ehssan Nazockdast[(1)] and Jeffrey F. Morris[(2)]

[(1)]Courant Institute of Mathematical Sciences, New York University; New York, NY 10012 USA

[(2)]Levich Institute and Department of Chemical Engineering, CUNY City College of New York; New York, NY 10031 USA

September 8, 2015

## Synopsis

Discrete particle simulations by Accelerated Stokesian Dynamics (ASD) and a microstructural theory are applied to study structure and the viscosity of hard-sphere Brownian suspensions in active microrheology (MR). The work considers moderate to dense suspensions, from near to far from equilibrium conditions. The microscopic theory explicitly considers many-body hydrodynamic interactions in active MR, and is compared with the results of the ASD simulations, which include detailed near and far field hydrodynamic interactions. We consider probe and bath particles which are spherical and of the same radius $a$. Two conditions of moving the probe sphere are considered: these apply constant force (CF) and constant velocity (CV), which approximately model magnetic bead and optical tweezer experiments, respectively. The structure is quantified using the probability distribution of colloidal particles around the probe, $P_{b|p}(\mathbf{r}) = ng(\mathbf{r})$, giving the probability of finding a bath particle centered at a vector position $\mathbf{r}$ relative to a moving probe particle instantaneously centered at the origin; $n$ is the bath particle number density, and is related to the suspension solid volume fraction, $\phi$, by $n = 3\phi/4\pi a^3$. The pair distribution function for the bath particles relative to the probe, $g(\mathbf{r})$, is computed as a solution to the pair Smoluchowski equation (SE) for $0.2 \leq \phi \leq 0.50$, and a range of Péclet numbers, describing the ratio of external force on the probe to thermal forces and defined $Pe_f = F^{ext}a/(k_bT)$ and $Pe_U = 6\pi\eta U^{ext}a^2/(k_bT)$ for CF and CV conditions, respectively. Results of simulation and theory demonstrate that a wake zone depleted of bath particles behind the moving probe forms at large Péclet numbers, while a boundary-layer accumulation develops upstream. The wake length saturates at $Pe_f \gg 1$ for CF while it continuously grows in CV. This contrast in behavior is related to the dispersion in the motion of the probe under CF conditions, while CV motion has no dispersion; the dispersion is a direct result of many-body hydrodynamic interactions. This effect is incorporated in the theory as a force-induced hydrodynamic diffusion flux in the pair SE. We also demonstrate that, despite this difference of structure in the two methods of moving the probe, the probability distribution of particles near the probe is primarily set by the Péclet number, for both CF and CV conditions, in agreement with dilute theories; as a consequence, similar values for apparent viscosity are found for the CF and CV conditions. Using the microscopic theory, the structural anisotropy and Brownian viscosity near equilibrium are shown to be quantitatively similar in both CF and CV motions, which is in contrast with the dilute theory which predict larger distortions and Brownian viscosities in CV, by a factor of two relative to CF microrheology. This difference relative to dilute theory arises due to the determining role of many-body interactions associated with the underlying equilibrium structure in the semi-dilute to concentrated regime.

# 1 Introduction

The field of microrheology (MR), as an alternative approach for measuring rheology of complex fluids at microscopic length scales, has undergone tremendous progress over the past two decades [MacKintosh & Schmidt, 1999; Waigh, 2005; Squires & Mason, 2010; Wirtz, 2009]. The motion of a probe particle (or particles), subjected to internal stimuli in passive MR or external stimuli in active MR, is measured and used to characterize the material properties in the vicinity of the probe(s). One appealing aspect of MR is that it allows analysis of the local microscopic structure of the material studied, facilitating investigation of microstructural heterogeneity [Valentine et al., 2001; Dasgupta & Weitz, 2005]. Another advantage of MR over bulk rheology is the significantly smaller volumes required, and its non-intrusive nature makes it ideal for rheological study of living systems [MacKintosh & Schmidt, 1999; Wirtz, 2009]. Other advantages include the large accessible frequency range in viscoelastic response [Willenbacher & Oelschlaeger, 2007], and direct measurement of interactions [Verma et al., 1998].

In a pioneering work, Mason & Weitz [Mason & Weitz, 1995] demonstrated that the linear viscoelastic response of the material can be extracted by monitoring the thermal fluctuations of a micron-sized probe particle embedded in the complex fluid using the generalized Stokes-Einstein relation (GSER). The theory developed by Mason & Weitz relates viscoelastic moduli to statistical measures of the probe motions such as mean square displacement or velocity auto-correlation [Squires & Mason, 2010]. Two major assumptions of the theory are *i*) the fluid must be near equilibrium and *ii*) the motion of the fluid about the probe should be inertialess, i.e. described by Stokes equations. A number of conditions can violate these two assumptions [Squires & Mason, 2010; Squires, 2008; Depuit & Squires, 2012a,b]. We focus on strongly nonlinear conditions, as these play a central role in the rheology of complex fluids. In many applications, complex fluids are driven far from equilibrium by externally applied stresses, resulting in nonlinear responses including shear thinning and thickening, as well as normal stress differences. Far from equilibrium, the GSER cannot be applied. Active microrheology is an alternative method to investigate the nonlinear response by externally moving the probe through the fluid using, for example, magnetic forces [Habdaas et al., 2004] or optical tweezers [Sriram et al., 2010]. Important questions addressed here are: To what degree can the results of active MR be compared against bulk shear rheology? What are the main differences between the two methods, and do those differences offer a new insight about the material? How is the behavior measured dependent upon the specific method of generating the probe particle motion? Specifically how do the measured structure and viscosity differ if the probe's velocity is controlled externally (CV) relative to the case where the probe is moved with a controlled external force (CF)?

The hard, or near-hard, sphere colloidal (Brownian particle) suspension is ideal for investigating these fundamental aspects of active MR, as a theoretical foundation of some completeness is already available for these systems, while the simulational tool of Stokesian Dynamics provides access to the structure and stresses to guide theoretical development for MR. Previous theoretical studies of MR for colloidal suspensions can be divided into two groups. The first group of works, by Brady and coworkers, are microscopic theories developed for dilute systems ($\phi \ll 1$) with interactions limited to pair level. For example, Squires & Brady [2005] studied dilute colloidal dispersions with only hard-sphere interactions (neglecting hydrodynamic interaction), and computed the distribution of bath particles around the probe as a function of the ratio of the external force to thermal forces (Péclet number, $Pe$). They observed that as the external force increases the distribution of colloidal particles around the probe becomes more anisotropic, with large accumulation of particles in front and a depleted zone or wake behind the probe. This work was extended to consider the effect of pair hydrodynamic interactions in dilute suspension [Khair & Brady, 2006], the role of probe particle's shape [Khair & Brady, 2008], relation between stress and force-induced diffusivity of the probe [Zia & Brady, 2012], and the case of having an oscillating external force/velocity [Swan et al., 2014]. The predictions of the dilute theories are in qualitative agreement with available experimental work [Sriram et al., 2010] and Brownian Dynamics simulations [Carpen & Brady, 2005]. The viscosity computed from the microstructure also shows qualitatively similar behavior as a function of Péclet number to shear viscosity although there are clear quantitative differences. A second group of theoretical studies has focused on dense suspensions near the jamming transition. These theories ignore hydrodynamic interactions between the probe and colloidal particles and use mode-coupling theory [Gazuz et al., 2009; Voigtmann & Fuchs, 2013] to predict the microstructure and from this extract the MR of colloidal suspensions. While these methods are powerful for studying very dense systems, due to the complete absence of hydrodynamic interactions, the hydrodynamic stress is not represented in these models. This limits the ability to accurately predict nonlinear behaviors, such as shear thickening in shear flow and the analogous force-thickening in active MR. There has also been a recent simulational study of active micrheology of soft-particle systems near jamming transition [Mohan et al., 2014]. The structure and rheology in this system are determined by elastohydrodynamic contact forces, which are not considered in our study.

In this study, we focus on the range of parameter space that has not been previously studied for MR, namely concentrated (up to $\phi = 0.50$) near-hard sphere colloidal suspensions and Péclet numbers ranging from near- ($Pe \ll 1$) to far-from-equilibrium ($Pe \gg 1$) conditions. We use Accelerated Stokesian Dynamics (ASD) simulations [Banchio & Brady, 2003] to study structure (pair distribution function, $g(\mathbf{r})$) and rheology (apparent viscosity) by modeling the active microrheology experiments. This is, to our knowledge, the first report on the discrete particle simulation of colloidal suspensions in active MR that includes detailed near- and far-field hydrodynamic interactions. Two conditions of regulating the motion of the probe are considered: the constant force (CF) approach, controlling the external force, $F^{ext}$, on the probe which mimics the experimental approach of pulling a magnetic bead probe [Habdaas et al., 2004; Choi et al., 2011]; and the constant velocity (CV) approach of setting the velocity, $U^{ext}$, of the probe, as is to a close approximation done by optical tweezers [Wilson et al., 2009; Sriram et al., 2010]. The Péclet number for CF and CV conditions are respectively defined as $Pe_f = (F^{ext}a)/k_bT$ and $Pe_U = (6\pi\eta U^{ext}a^2)/k_bT$, where $a$ is the radius of the probe, $\eta$ is the viscosity of the fluid medium, and $k_bT$ is the thermal energy. The cases of CF and CV microrheology may be regarded, respectively, as the analogs of bulk rheometry in controlled stress and controlled rate of strain modes, which typically yield similar results. In active MR it is not certain that the two methods are equivalent.

We use sampling from ASD simulations to show that the structural anisotropies induced by the motion of the probe in CF and CV are distinctly different at $Pe_{f,U} \gg 1$: In CV the depleted zone of bath particles behind the probe monotonically extends with increasing the Péclet number (or external velocity). By contrast when moving the probe with constant force, the length of depleted zone (and other structural features away from the probe) saturate with $Pe_f$. We demonstrate that this saturation is a result of the dispersive motion of the probe in CF conditions, while in CV the probe's trajectory is a straight line. We show that the source of the dispersion in CF is the hydrodynamic interactions of the probe with its neighboring bath particles. This is the first study that takes into account the effect of hydrodynamic dispersion on the structure and viscosity of colloidal suspensions in active microrheology; the previous theories ignore this effect by either assuming the suspension is dilute and the interactions are limited to pair particles [Squires & Brady, 2005; Khair & Brady, 2006; Swan & Zia, 2013; Swan et al., 2014], or the HI is entirely ignored [Gazuz et al., 2009; Voigtmann & Fuchs, 2013].

We complement the simulation results with a microscopic theory developed by modifying our theoretical framework for shear flows [Nazockdast & Morris, 2012b,a, 2013] for active microrheology. We consider both conditions of CF and CV. The theory is based on computation of the probe-bath particle pair correlation, described through a pair distribution function, $g(\mathbf{r})$, determined as a solution to the appropriate form of the pair Smoluchowski equation (PSE). This theory captures through probabilistic means of incorporating many-body effects the interactions that lead to dense-packed structural features, e.g. next-nearest neighbor rings. The dispersion of the probe's trajectory in CF conditions is incorporated in our theory by adding a diffusive flux term in the PSE, which points to the other unique aspect of this theory. The force-induced diffusion coefficient is modeled by extending the equilibrium relationship between diffusion and particle pressure to non-equilibrium conditions. The predictions of the theory are in good agreement with the ASD results over the range of $Pe$ and $\phi$ studied, although discrepancies will be noted.

Our simulation results and predictions of theory show that the boundary layer structure of the density of the particles formed very near the probe at $Pe \gg 1$ scales similarly with $Pe$ for CF and CV (unlike the structure away from the probe). We specifically show that the pair distribution function is determined by the net external force on the probe, irrespective of the volume fraction and method of moving the probe. This finding is in agreement with the predictions of the dilute theories. Due to this similarity of near-contact structure for CF and CV conditions, similar values of $g$ were computed for apparent viscosity at $Pe \gg 1$ from theory and simulation for the two forcing conditions.

The near-equilibrium structure and rheology is only studied theoretically, due to the poor sampling statistics of ASD simulations at $Pe < 1$. The predicted structural distortion at $Pe \ll 1$ for CF and CV are similar in magnitude and form, resulting in similar values for their respective viscosities. This is not in agreement with predictions of dilute theories which predict structural distortions and Brownian viscosities twice as large for CV conditions. We show that this difference arises because of the near-contact excluded volume interactions which integrate to an effective mean force from the bath particles in PSE; this effect is completely absent in the dilute theories where the pair distribution function is near unity everywhere. These many-body interactions result in a significantly faster decay of distorted structure away from the probe in the present theory as compared to the predictions of the dilute theories.

## 2 Simulation of active microrheology

The Accelerated Stokesian Dynamic (ASD) simulation method is applied to compute the dynamics of $N$ Brownian hard-sphere particles interacting hydrodynamically. The sampled positions of the spheres are then used to compute the pair distribution function, $g(\mathbf{r})$, and make direct comparison with predictions of the theory. The details of the conventional Stokesian Dynamics (SD) and the more computationally efficient implementation ASD are given elsewhere [Sierou & Brady, 2001; Banchio & Brady, 2003]. A brief description of the method and its implementation for active microrheology are discussed here.

Stokesian Dynamics computes the dynamics of particles immersed in an incompressible Newtonian fluid with negligible inertial effects, i.e. at vanishing Reynolds number. The hydrodynamic interaction between the particles is governed by Stokes equations and the motion of the particles is determined through Langevin equation

$$\mathbf{m} \cdot \frac{d\mathbf{U}}{dt} = \mathbf{F}^H + \mathbf{F}^B + \mathbf{F}^P + \mathbf{F}^{ext}, \quad (1)$$

where in a system composed of $N$ particles, $\mathbf{m}$ is the generalized mass/moment-of-inertia $6N \times 6N$ matrix, $\mathbf{U}$ is the generalized translational/rotational velocity vector of particles of size $6N$, and $\mathbf{F}^H, \mathbf{F}^B, \mathbf{F}^P$ are generalized hydrodynamic, Brownian, interparticle force/torque vectors of size $6N$, respectively. The Langevin equation is then integrated in time steps large enough that the momentum of the particles is relaxed ($\mathbf{m} \cdot d\mathbf{U}/dt \approx 0$) $\Delta t \gg \tau^m = m/6\pi\eta a$, but small enough that the configuration of the particles is not substantially changed due to Brownian diffusion or external flows; the result is the over-damped Langevin equation for Brownian particles. This equation is again integrated in time with $\Delta t \gg \tau^m$, to obtain the evolution equation for the positions and alignments of the particles [Banchio & Brady, 2003].

In active microrheology, a force is also applied externally to drag the probe through the bath particles which is represented with $\mathbf{F}^{ext}$. Note that $\mathbf{F}^{ext}$ is zero for all particles except the probe. In the absence of externally imposed flows, we have $\mathbf{F}^H = \mathbf{R}^{FU} \cdot \mathbf{U}$ for hydrodynamic and interparticle forces where $\mathbf{R}^{FU}$ is the grand resistance tensor of size $6N \times 6N$ which relates force to velocity of the particles and only depends on the positions of the particles. For particles near contact, the resistance matrix is computed using a pairwise additive lubrication resistance formulation and when particles are well-separated, a pairwise additive mobility formulation is adopted, with the mobility matrix inverted to obtain the resistance matrix. The Brownian force is random with zero time average, and strength given according to the fluctuation-dissipation theorem by $\langle \mathbf{F}^B(0)\mathbf{F}^B(t)\rangle = 2k_b T \mathbf{R}^{FU} \delta(t)$. Once the resistance matrix is known from the configuration of particles, the velocity of the particles including the probe can be computed as a solution to (1). Afterwards, (1) is integrated in time to update the positions of particles. Periodic boundary conditions are considered in all three directions.

The implementation for constant force on the probe (CF) is straight forward; the forces, including $\mathbf{F}^{ext}$ applied only on the probe, are explicitly known at each time step, with $\mathbf{F}^B$ drawn from a random distribution. The velocities are calculated as a solution to the set of linear equations defined by (1). Under precise CV conditions, $\mathbf{F}^H$ (again, on the probe alone) as well as $\mathbf{U}^H$ and $\mathbf{U}^B$ are unknowns, which presents difficulties that were avoided by numerically replicating what is done in laser tweezer experiments [Sriram et al., 2010]. Instead of pulling the probe with a truly constant velocity $\mathbf{U}^{ext}$, we simulated the reverse case of fixing the probe while the bath particles are all in the external flow of $\mathbf{U}^\infty = -\mathbf{U}^{ext}$. A spring force proportional to deviation from its initial position was applied to the probe, $\mathbf{F}_1^{ext}(t) = -K(\mathbf{x}_1(t) - \mathbf{x}_1(0))$ where $K$ is the spring constant. As the simulation begins the probe is displaced from its initial position in the direction of the external flow until the spring force, in average, balances the force exerted by the bath particles. Due to fluctuations in microstructure, the probe position fluctuates around its average value, which is slightly displaced in the flow direction, $|\Delta\mathbf{x}|/a \sim 6\pi\eta U^{ext}/K$. This slight fluctuation may prove important relative to a precisely fixed velocity, but is actually a better physical mimic of experimental conditions. As an example of the behavior, in figure 1 we illustrate trajectories of the probe under CV conditions at $Pe_U = \frac{6\pi\eta_0 U^{ext} a^2}{k_b T} = 10$ and $\phi = 0.40$. These are obtained from simulations with different initial conditions. The trajectories are projected to the $x - z$ plane where $z$ is the direction of the probe's motion; note that $x$ and $y$ are equivalent directions owing to axisymmetry about the probe's mean motion. As pointed out, the probe undergoes a mean displacement in the direction of the flow of a magnitude $\langle |\mathbf{x}|/a \rangle \approx 0.6$, while the displacement in other directions is zero in average. Viscosity is computed as $\eta_U = -\langle \mathbf{F}^{ext} \rangle / (6\pi U^{ext} a)$ where $\langle \mathbf{F}^{ext} \rangle = -K (\langle \mathbf{x}_1 \rangle - \mathbf{x}_1(0))$ is the ensemble average of the external force, taken after the displacement of the probe reaches a statistically steady behavior. The spring constant was chosen proportional to $Pe_U$ at $Pe_U \geq 1$ such that probe remains within a radius of $0.2a$ from its initial position for the large majority of the time.

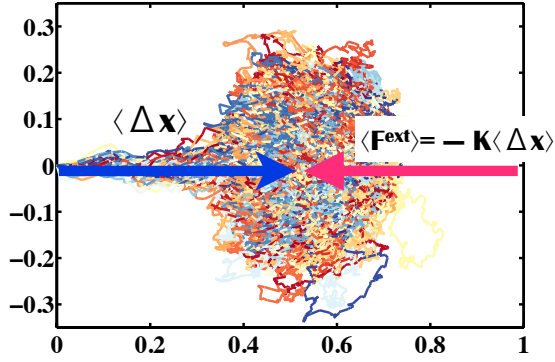

**Figure 1:** *The trajectories of the probe particle in CV condition from simulations with different initial conditions at $Pe_U = 10$ and $\phi = 0.40$, projected to $x - z$ plane, where $z$ is the direction of the probe motion.*

The number of configurations needed to achieve a statistically meaningful structure in active microrheology is significantly larger than in shear flow as the structure is measured only around the probe. Thus, we took statistics from 30 to 60 simulations over an extended period of motion for each $Pe$ and $\phi$. The structural distortions from isotropy induced by pulling the probe through the suspension near equilibrium ($Pe \ll 1$) are proportional to Péclet, and this was not found to be feasible to access by simulation with our available resources. To illustrate the problem, our simulation results from 60 simulation runs at $Pe_f = 1$, $\phi = 0.40$ yield $(g^{max} - g^{eq})/g^{eq} < 0.07$ where $g^{max}$ is the maximum value of $g(\mathbf{r})$. For smaller $Pe_f$, the magnitude of anisotropy $(g^{max} - g^{eq})$ will be smaller, and we find that the number of simulations performed in this study is not large enough to obtain a statistically reliable measure of $g - g_{eq}(\mathbf{r})$ at $Pe_f < 0.5$. Thus we present ASD results for $Pe_f \geq 1$ and compare with the theoretical results, while for study $Pe_f < 1$ only the theory is used (see

§6.1).

The far-from-equilibrium condition, i.e. $Pe \gg 1$, is characterized by the formation of a wake region depleted of bath particles behind the probe. This wake can extend to many particle diameters, resulting in a complication for the simulation technique applied here, as the probe may interact with the wake formed by its periodic image. To avoid this, simulations with $N = 2048$ particles were performed for these conditions. Nevertheless, under CV conditions we find the wake to be especially pronounced, and in fact it extends over the entire length of the simulation box (over ten particle diameters) at $Pe_U \geq 25$.

# 3 Simulated microstructure and trajectories

In this section, we present a limited set of results, considering only the simulated probe trajectories and the microstructure at $Pe \geq 1$. These are obtained from ASD simulation implemented as described in the prior section. The microrheology viscosity results determined from simulation (as described in §5) will be presented in §6 and 7, where they will be compared with the theory. The theory is formulated in the following §4. The results of this section serve to emphasize the key difference between CF and CV conditions. In the latter where its velocity is fixed, the probe travels in an essentially straight line through the suspension, with only the slight fluctuations allowed by the spring force. In contrast to this, for a fixed force, the probe changes its direction and its speed in response to the heterogeneity of the local structure of the suspension bath. We begin by showing that this difference in the trajectory of the probe results in distinct large-scale structural differences when $Pe \gg 1$. Here, we recall that $Pe$ denotes the ratio of the external force applied on the probe to thermal forces and will be specified to $Pe_f = F^{ext}a/(k_bT)$ and $Pe_U = 6\pi\eta U^{ext}a^2/(k_bT)$ for CF and CV conditions, respectively, where needed for clarity. Here, $\eta$ is the viscosity of the suspending fluid.

The microstructure is described by the distribution of number density of bath particles around the probe, $P(\mathbf{r}) = ng(\mathbf{r})$ where $n = N/V$ is the average number density and $g(\mathbf{r})d\mathbf{r}$ is thus the normalized likelihood of observing a bath particle between $\mathbf{r}$ and $\mathbf{r} + d\mathbf{r}$; note that this normalization implies $g \to 1$ at large $r = |\mathbf{r}|$ where the influence of the probe on the bath particle distribution vanishes.

Figures 2(a) to 2(c) illustrate $g(\mathbf{r})$ from sampling the simulated configurations of the bath particles around a probe (whose center serves as the origin) at $\phi = 40$ and $Pe_U = 10, 25$ and 100. At $Pe_U = 10$ a boundary layer structure is formed with a zone of large $g(\mathbf{r})$ near contact in the direction of motion and a zone depleted of bath particles (wake zone) behind the probe. The length of the wake zone increases with the flow strength, and at $Pe_U = 25$ the wake spans the entire simulation box. Consequently, the probe interacts with its periodic image which results in a depleted streak in front of the probe at $Pe_U = 25$ and $Pe_U = 100$.

The sampled structures for CF condition at $\phi = 0.40$ and $Pe_f = 100, 400$, and 1000 are shown in figures 2(d) to 2(f); recall $Pe_f = \frac{F^{ext}a}{k_bT}$. A boundary layer structure, similar to that found in CV conditions, is observed at $Pe_f = 100$. The near contact values of $g(\mathbf{r})$ in front of the probe (shown as the maximum value on the color bars) monotonically increase from $Pe_f = 100$ to 1000. The length of the wake zone increases from $Pe_f = 100$ to $Pe_f = 400$. Aside from the increase of contact values with Péclet number, the structure (including the length of the wake zone) remains unchanged from $Pe_f = 400$ to 1000. This saturation of the structure far from contact, with the most noticeable feature being the wake length, marks the essential difference we have observed between CF and CV microrheology (MR) at $Pe \gg 1$.

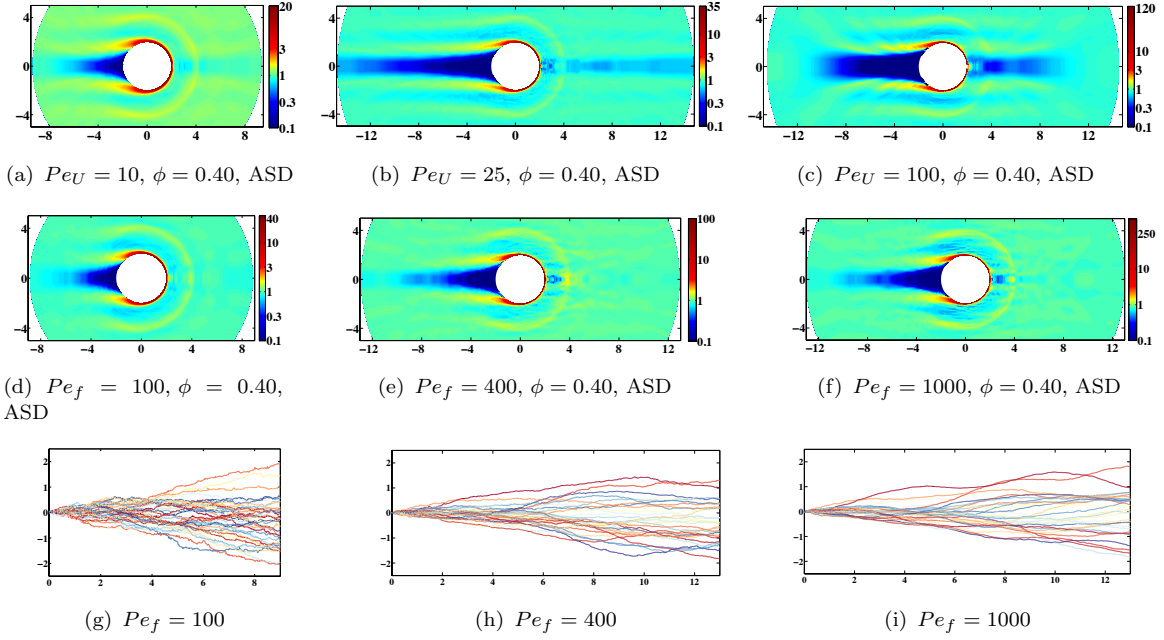

**Figure 2:** *(a-f) The computed pair microstructure, $g(\mathbf{r})$, of bath particles around the probe obtained from ASD simulations; (a-c) constant velocity mode of motion at $\phi = 0.40$ and $Pe_U = 10, 25,$ and $100$, respectively; (d-f) constant force mode at $\phi = 0.40$ and $Pe_f = 10, 100,$ and $1000$, respectively. The maximum value in each color bar indicates the scale of the near-contact pair probability. The lower limit of the color bars representing $\mathcal{O}(1)$ features of $g(\mathbf{r})$ are similar for all Péclet numbers to facilitate comparison. The panels 2(g) to 2(i) illustrate simulated trajectories of the probe, projected onto the x-z plane worth z the direction of motion, under CF conditions at $\phi = 0.40$ and $Pe_f = 100, 400$ and $1000$, respectively. A dispersion around the average trajectory of the probe is observed, with similar dispersion for $Pe_f = 400$ and $1000$.*

In figures 2(g) to 2(i) we present the trajectories of the probe particle at $Pe_f = 100$, 400 and 1000, corresponding to the microstructures in figures 2(d) to 2(f). The physical origin of the contrast between CV and CF structure at $Pe \gg 1$ is seen in these trajectories, which clearly show a hydrodynamic dispersion around the average motion of the probe. The observed dispersion at $Pe_f \gg 1$ is a consequence of hydrodynamic interactions (HI), and does not have thermal origin. This becomes evident by noting that the variation of the dispersion radius with traveling distance remains almost unchanged as the conditions are changed from $Pe_f = 400$ to $Pe_f = 1000$, i.e. the dispersion is proportional to the distance traveled, $U^{ext}t$, regardless of the time taken to travel the distance. The wake zone is filled in by particles owing to the hydrodynamic dispersion, reaching uniform number density ($g \to 1$) at sufficient distance from the probe: this distance saturates at large $Pe$ because the dispersion itself saturates. The observations show that the length of the wake zone at $Pe \gg 1$ is roughly the axial distance traveled in order for the probe motion to show a significant probability of lateral displacement by its own radius through the dispersion process; the length of the wake zone eventually saturates as $Pe \to \infty$ and for $\phi = 0.40$, the wake extent is approximately $10a$. For CV conditions, there is only limited hydrodynamic dispersion of the bath particles and not of the probe, so the wake fills in primarily by thermal diffusion of the bath particles, which is much slower than the translation of the probe at large $Pe_U$, leading to the very long wakes at fixed probe velocity.

Because the dispersion of the probe's trajectory is induced by many-body HI, it increases as $\phi$ increases, resulting in reduction of the length of the wake. We demonstrate this by showing $g(\mathbf{r})$ and probe trajectories at $Pe_f = 100$ and different volume fractions in figure 3.

A quantitative measure of the dispersion around the average motion can be obtained by computing the mean-square displacement (MSD) in the plane perpendicular to the direction of motion. In figure 4(a) we present the MSD as a function of the distance travelled, denoted $U^{ext}t$, by the probe at $\phi = 0.40$ and $Pe_f = 100$, 400 and 1000; this is taken for convenience along one coordinate direction, $y$, but the dynamics is axisymmetric around the probe's motion (along $z$), and the dispersion in $x$ is equivalent. Note that for distances traveled smaller than the particle size, the extent of dispersion is larger at the smaller values of $Pe_f$. For $U^{ext}t > 5$, however, we see convergence of the displacements at all $Pe_f$ to a single curve. The observed behavior can be explained by noting that the dispersion at small $U^{ext}t$ is influenced by Brownian motion, which at a fixed velocity is stronger at smaller $Pe_f$. The force-induced dispersion is a result of the probe interacting with a sequence of different configurations of neighboring particles as it moves through the suspension, and when the probe has interacted with many bath particles ($U^{ext}t > 10a$), the force-induced dispersion dominates the Brownian diffusion such that the extent of dispersion shown as MSD *vs* $U^{ext}t$ becomes independent of $Pe$. We plot the MSD for different volume fractions in figure 4(b), showing the increase of dispersion at larger $\phi$.

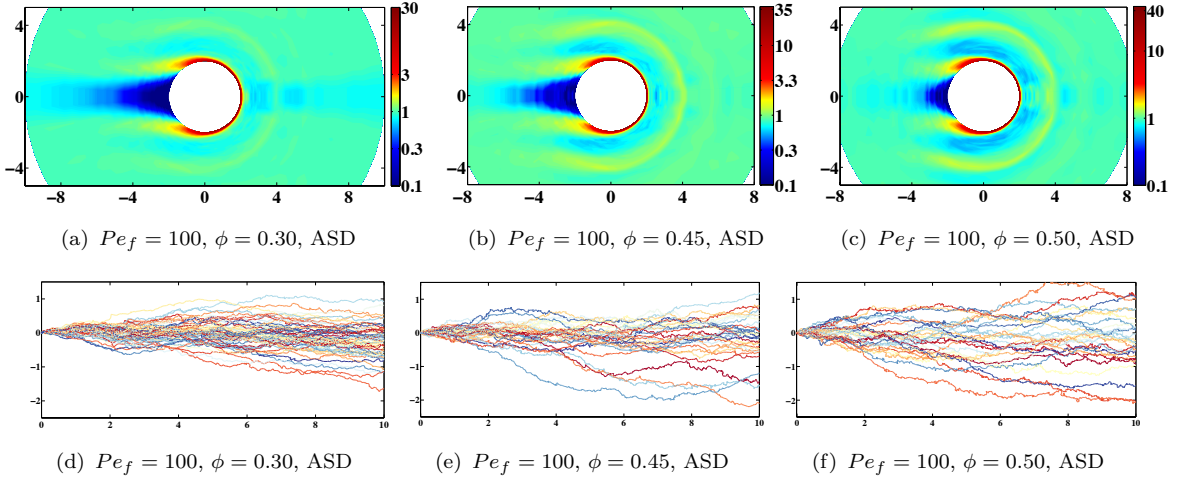

**Figure 3:** *(a-c) The sampled $g(\mathbf{r})$ from ASD simulation under CF conditions at $Pe_f = 100$, for $\phi = 0.30$, $0.45$ and $0.50$. The wake zone is reduced in size with increasing $\phi$. (d-f) The trajectories of the probe for individual simulations for the same conditions as figures (a-c), respectively. The dispersion around the average motion increases with volume fraction resulting in a reduction of the wake zone.*

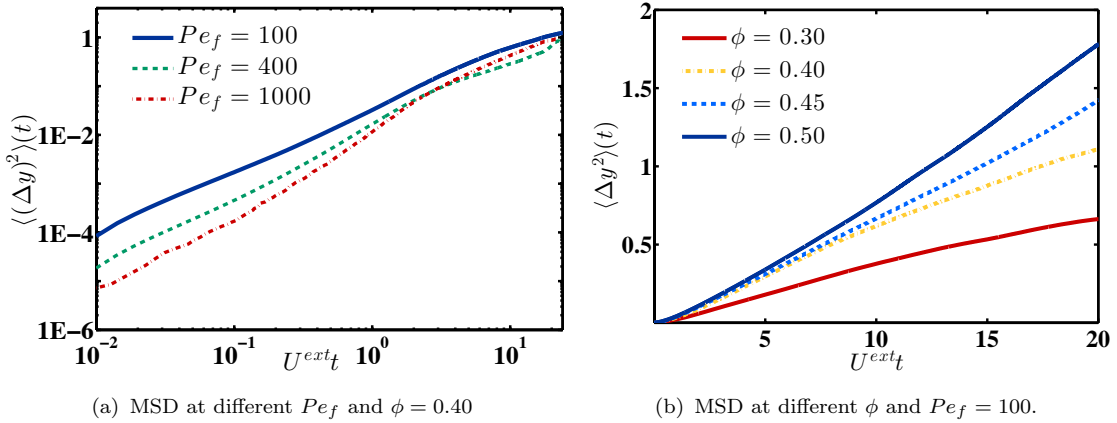

**Figure 4:** *The mean squared displacement (MSD) of the probe particle motion, with distances scaled by the particle radius $a$. (a) MSD in the directions orthogonal to the direction of motion at $\phi = 0.40$ and $Pe_f = 100$, $400$, and $1000$; (b) MSD at $Pe_f = 100$ and $\phi = 0.30$, $0.40$, $0.45$ and $0.50$.*

We note that the force-induced dispersion has not been included in previous theories for active microrheology of colloidal dispersions. The previous theories have either neglected many-body interactions by assuming the suspension is dilute [Squires & Brady, 2005; Khair & Brady, 2006; Swan & Zia, 2013], or only consider steric interactions and simply discard hydrodynamic interactions [Gazuz *et al.*, 2009; Gnann *et al.*, 2011; Voigtmann & Fuchs, 2013]. In the next section we present our microscopic theory which incorporates the force-induced dispersion as well as the average velocity induced by many-body particle and hydrodynamic interactions. These features are found to be necessary to allow the theory developed here to reasonably predict the microstructure and viscosity up to $\phi = 0.40$ and $Pe \gg 1$.

## 4 Microstructural theory formulation

Here we develop the theoretical description, which follows the approach used in our theory developed for the microstructure of sheared suspensions [Nazockdast & Morris, 2012*b*, 2013]. We describe first the case of a constant velocity probe, and then the constant force case. Results of the theory will be presented in §6 and §7, with comparison to ASD results at $Pe \geq 1$.

### 4.1 Constant velocity probe

We consider $N$ identical spherical particles dispersed in an incompressible Newtonian fluid. One of these particles, labeled 1, is defined to be the probe particle that is pulled through the bath of particles with a constant velocity (CV), $\mathbf{U}_1$. The ultimate goal is to compute the force exerted on the probe by the bath particles and fluid. The motion of the probe particle through the suspension changes the spatial distribution of force- and torque-free bath particles around it. The resistance to motion of the probe depends on this distribution, and consequently the force and suspension microstructure are coupled. The $N$-body configurational probability distribution function, $P_N(\mathbf{x}_1, \mathbf{x}_2, ..., \mathbf{x}_N)$, satisfies the conservation equation

$$\frac{\partial P_N}{\partial t} + \sum_{\alpha=1}^{N} \nabla_\alpha \cdot \mathbf{j}_\alpha = 0, \qquad (2)$$

where $\mathbf{j}_\alpha$ is the probability flux associated with particle $\alpha$. When pulling the probe with a *fixed velocity*, the probability flux associated with the probe is purely convective and is expressed as $\mathbf{j}_1 = \mathbf{U}_1 P_N$. The remaining particles move freely in response to Brownian, interparticle and hydrodynamic forces, with the flux for $\alpha \neq 1$ expressed as

$$\mathbf{j}_\alpha = \mathbf{U}_\alpha^H P_N + \left( \sum_{\beta=1}^{\beta=N} \mathbf{M}_{\alpha\beta}^{UF} \cdot \mathbf{F}_\beta^P \right) P_N - \sum_{\beta=1}^{\beta=N} \mathbf{D}_{\alpha\beta} \cdot \nabla_\beta P_N. \tag{3}$$

Here $\mathbf{U}_\alpha^H$ is the hydrodynamic velocity of particle $\alpha$ induced by motion of the probe particle through its hydrodynamic interaction with bath particles, $\mathbf{M}_{\alpha\beta}^{UF}$ is the mobility of particle $\alpha$ due to a force on particle $\beta$, and $\mathbf{D}_{\alpha\beta} = k_b T \mathbf{M}_{\alpha\beta}^{UF}$ is the Brownian diffusion tensor. The velocities of particle $\alpha$ induced by interparticle and Brownian forces applied on the rest of the particles are given, respectively, by

$$\mathbf{U}_\alpha^P = \sum_{\beta=1}^{\beta=N} \mathbf{M}_{\alpha\beta}^{UF} \cdot \mathbf{F}_\beta^P \quad \text{and} \quad \mathbf{U}_\alpha^B = -\sum_{\beta=1}^{\beta=N} \mathbf{D}_{\alpha\beta} \cdot \nabla_\beta P_N / P_N.$$

To allow progress, we reduce $P_N$ to the distribution of number density of bath particles around the probe, $P(\mathbf{r}) = ng(\mathbf{r})$ where $n = N/V$ is the average number density and $g(\mathbf{r})$ is the pair distribution function for the bath-probe pair. To reduce to an equation for $g(\mathbf{r})$, (2) is integrated over all possible configurations of $N-2$ particles keeping particle 1 and one bath particle (which we refer to as particle 2) fixed (meaning of known position, but mobile). This yields

$$\frac{\partial g}{\partial t} + \nabla \cdot \langle \mathbf{j}_2 - \mathbf{j}_1 \rangle_2 = 0. \tag{4}$$

The average of a function $A(\mathbf{r})$ at a fixed separation is denoted $\langle A \rangle_2(\mathbf{r})$, with the averaging performed over all possible configurations of the other $N-2$ bath particles. We make a mean-field approximation, assuming that the particle pair of interest is immersed in a field that contains the average effect of the other particles. As a consequence, we neglect the convective flux associated with fluctuations of force and mobility:

$$\langle \mathbf{M} \cdot \mathbf{F} \rangle_2 = \langle \mathbf{M} \rangle_2 \cdot \langle \mathbf{F} \rangle_2. \tag{5}$$

Applying the mean-field approximation to (4), the average velocities of particles 1 and 2 are related to the average hydrodynamic forces on each particle as follows:

$$\begin{bmatrix} \langle \mathbf{F}_1^H \rangle_2 \\ \langle \mathbf{F}_2^H \rangle_2 \end{bmatrix} = \begin{bmatrix} \mathbf{R}_{11} & \mathbf{R}_{12} \\ \mathbf{R}_{21} & \mathbf{R}_{22} \end{bmatrix} \cdot \begin{bmatrix} -\mathbf{U}_1 \\ -\langle \mathbf{U}_2 \rangle_2 \end{bmatrix} \tag{6}$$

where

$$\begin{bmatrix} \mathbf{R}_{11} & \mathbf{R}_{12} \\ \mathbf{R}_{21} & \mathbf{R}_{22} \end{bmatrix} = \begin{bmatrix} \langle \mathbf{M}_{11}^{UF} \rangle_2 & \langle \mathbf{M}_{12}^{UF} \rangle_2 \\ \langle \mathbf{M}_{21}^{UF} \rangle_2 & \langle \mathbf{M}_{22}^{UF} \rangle_2 \end{bmatrix}^{-1}. \tag{7}$$

Note that the formulation is written for hydrodynamic forces applied from the fluid to the particles, resulting in a negative sign on velocities in (6). Since the probe velocity is assumed fixed, $\langle \mathbf{U}_1^B \rangle_2 = 0$ and the Brownian force on particle 2 from the bath particles is

$$\langle \mathbf{F}_2^B \rangle_2 = -k_b T \nabla g(\mathbf{r})/g(\mathbf{r}). \tag{8}$$

Recall that the bath particles are force- and torque-free on timescales longer than the momentum relaxation time, $\langle \mathbf{F}_2 \rangle_2 = \langle \mathbf{F}_2^H \rangle_2 + \langle \mathbf{F}_2^B \rangle_2 + \langle \mathbf{F}_2^P \rangle_2 = 0$, and incorporating (6) and (8) to this condition for the bath particles, $\langle \mathbf{U}_2 \rangle_2$ is written as

$$\langle \mathbf{U}_2 \rangle_2 = (\mathbf{R}_{22})^{-1} \cdot \left[ -\mathbf{R}^{21} \cdot \mathbf{U}_1 + \langle \mathbf{F}_2^P \rangle_2 - k_b T \nabla g(\mathbf{r})/g(\mathbf{r}) \right]. \tag{9}$$

Taking $\langle \mathbf{j}_2 - \mathbf{j}_1 \rangle_2 = (\langle \mathbf{U}_2 \rangle_2 - \mathbf{U}_1) g(\mathbf{r})$, (4) in steady state is modified to the pair SE for CV conditions,

$$\nabla_\mathbf{r} \cdot \left[ \mathcal{C} \cdot \mathbf{U}_1 g(\mathbf{r}) + \mathbf{M} \cdot \langle \mathbf{F}_2^P \rangle_2 g(\mathbf{r}) - k_b T \mathbf{M} \cdot \nabla_\mathbf{r} g(\mathbf{r}) \right] = 0, \tag{10}$$

where $\mathcal{C} = -(\mathbf{R}_{22})^{-1} \cdot \mathbf{R}_{21} - \mathbf{I}$ and $\mathbf{M} = (\mathbf{R}_{22})^{-1}$. Quantities are made dimensionless as follows:

$$\mathbf{r} \sim a, \quad \langle \mathbf{M}_{\alpha\beta} \rangle_2 \sim M_0, \quad \mathbf{R}_{\alpha\beta} \sim M_0^{-1}, \quad \mathbf{U}_1 \sim U^{ext}, \quad \langle \mathbf{F}^P \rangle_2 \sim k_b T/a,$$

where $M_0 = \dfrac{1}{6\pi\eta a}$ is the single particle mobility. The pair SE in dimensionless form is

$$\nabla \cdot \left[ \mathcal{C} \cdot \mathbf{U}_1 g(\mathbf{r}) + Pe_U^{-1} \mathbf{M} \cdot \left( \langle \mathbf{F}_2^P \rangle_2 g(\mathbf{r}) - \nabla g(\mathbf{r}) \right) \right] = 0, \tag{11}$$

and we recall that

$$Pe_U = \frac{U^{ext} a}{k_b T M_0} = \frac{6\pi \eta U^{ext} a^2}{k_b T}. \tag{12}$$

The boundary conditions for (11) are

$$\mathbf{j_r} = 0 \quad \text{at} \quad r = 2, \tag{13a}$$
$$g \to 1 \quad \text{as} \quad r \to \infty. \tag{13b}$$

The first boundary condition imposes zero radial flux at contact due to excluded volume. In order to solve (11), the conditional averages in the equation including average pair mobility and interparticle force should be formulated in terms of $\mathbf{r}$ and $g(\mathbf{r})$. The formulations are given in §4.3.

## 4.2 Constant force probe

We now consider the case where the probe particle is pulled through the bath with a constant force (CF), $\mathbf{F}^{ext}$, with the force taken along the $z$ axis. The dispersive flux induced by velocity fluctuations of the probe and bath particles is the fundamental difference of CF and CV active microrheology. In a CF experiment, unlike when the probe is dragged with a constant velocity, the probe can change direction and speed. As a result, the probe trajectory is not always in the direction of the applied force, resulting in the noted dispersion. If the probe is pulled at $Pe_f = \dfrac{F^{ext}a}{k_bT} \gg 1$, thermal fluctuations become negligible with respect to the force-induced fluctuations in setting the long-time dispersion; thus it is essential to include dispersion in the probe motion when formulating the structure problem for active mirorheology with CF at $Pe_f \gg 1$.

The CF formulation follows the CV formulation until the relation (4). The probability fluxes are calculated based on a mobility formulation for CF conditions. The hydrodynamic velocities of the probe and particle 2 are expressed as

$$\begin{bmatrix} \langle \mathbf{U}_1^H \rangle_2 \\ \langle \mathbf{U}_2^H \rangle_2 \end{bmatrix} = \begin{bmatrix} \langle \mathbf{M}_{11}^{UF} \rangle_2 & \langle \mathbf{M}_{12}^{UF} \rangle_2 \\ \langle \mathbf{M}_{21}^{UF} \rangle_2 & \langle \mathbf{M}_{22}^{UF} \rangle_2 \end{bmatrix} \cdot \begin{bmatrix} \mathbf{F}^{ext} \\ 0, \end{bmatrix}, \tag{14}$$

which gives

$$\mathbf{U}^H = \langle \mathbf{U}_2^H \rangle_2 - \langle \mathbf{U}_1^H \rangle_2 = - \left( \langle \mathbf{M}_{11}^{UF} \rangle_2 - \langle \mathbf{M}_{21}^{UF} \rangle_2 \right) \cdot \mathbf{F}^{ext}. \tag{15}$$

Note that the bath particles are force-free, i.e. $\mathbf{F}_2^{ext} = 0$ and $\mathbf{F}^{ext}$ in (14) is applied externally; as a result, the hydrodynamic force applied on particles from the fluid to balance the external force is $-\mathbf{F}^{ext}$, and in (14), unlike (6), there is no negative sign applied to the velocity components. The relative velocity associated with interparticle forces can be computed using a relation similar to (14), given by

$$\mathbf{U}^P = \langle \mathbf{U}_2^P \rangle_2 - \langle \mathbf{U}_1^P \rangle_2 = (\langle \mathbf{M}_{22} \rangle_2 - \langle \mathbf{M}_{12} \rangle_2) \cdot \langle \mathbf{F}_2^P \rangle_2 + (\langle \mathbf{M}_{21} \rangle_2 - \langle \mathbf{M}_{11} \rangle_2) \cdot \langle \mathbf{F}_1^P \rangle_2, \tag{16}$$

where the $UF$ superscript of the mobility tensor $\mathbf{M}$ is dropped for convenience.

We model the flux associated with the flow-induced dispersion as a diffusion. The relative flux due to velocity fluctuations is therefore written as

$$\mathbf{j}_2^D - \mathbf{j}_1^D = -(\mathbf{D}_{22} - \mathbf{D}_{12} + \mathbf{D}_{11} - \mathbf{D}_{21}) \cdot \nabla g(\mathbf{r}), \tag{17}$$

where

$$\mathbf{D}_{22} = k_b T \langle \mathbf{M}_{22} \rangle_2, \tag{18a}$$

$$\mathbf{D}_{12} = k_b T \langle \mathbf{M}_{12} \rangle_2, \tag{18b}$$

$$\mathbf{D}_{11} = k_b T \langle \mathbf{M}_{11} \rangle_2 + \mathbf{D}_{11}^f, \tag{18c}$$

$$\mathbf{D}_{21} = k_b T \langle \mathbf{M}_{21} \rangle_2 + \mathbf{D}_{21}^f, \tag{18d}$$

where the $D$ superscript on $\mathbf{j}$ in (17) is indicative of the diffusion flux. The force-induced dispersion only appears on $\mathbf{D}_{21}$ and $\mathbf{D}_{11}$, since only interparticle and thermal forces (and no external force) act on particle 2 ($\mathbf{D}_{22}^f$ and $\mathbf{D}_{12}^f$ are zero); recall that $\mathbf{D}_{\alpha\beta}$ is the diffusion of particle $\alpha$ induced by the force on particle $\beta$. The diffusion induced by Brownian forces scales as $k_b T M_0$ where $k_b T/a$ is the thermal force, and $M_0$ is the single particle mobility. When $Pe_f = \dfrac{F^{ext} a}{k_b T} \gg 1$, the diffusion of the probe is expected to scale with $F^{ext}$ instead of $k_b T/a$, i.e. $\mathbf{D}^f \sim M_0 F^{ext} a$. Based on the given relations, the CF pair SE reduces to

$$\nabla_{\mathbf{r}} \cdot \left[ \left(\mathbf{U}^H(\mathbf{r}) + \mathbf{U}^P(\mathbf{r})\right) g(\mathbf{r}) - \left(\mathbf{D}^B(\mathbf{r}) + \mathbf{D}^f(\mathbf{r})\right) \cdot \nabla_{\mathbf{r}} g(\mathbf{r}) \right] = 0, \tag{19}$$

with

$$\mathbf{D}^B = k_b T \left( \langle \mathbf{M}_{11} \rangle_2 - \langle \mathbf{M}_{21} \rangle_2 + \langle \mathbf{M}_{22} \rangle_2 - \langle \mathbf{M}_{12} \rangle_2 \right), \tag{20a}$$

$$\mathbf{D}^f(\mathbf{r}) = \mathbf{D}_{11}^f(\mathbf{r}) - \mathbf{D}_{21}^f(\mathbf{r}), \tag{20b}$$

where $\mathbf{D}^B(\mathbf{r})$ and $\mathbf{D}^f(\mathbf{r})$ are relative diffusivities induced by Brownian and external forces. Quantities are made dimensionless using the scalings

$$\mathbf{r} \sim a, \quad \mathbf{U}^H \sim M_0 F^{ext}, \quad \mathbf{U}^P \sim k_b T M_0/a, \quad \mathbf{D}^B \sim k_b T M_0, \quad \mathbf{D}^f \sim M_0 F^{ext} a.$$

The dimensionless form of the pair SE is

$$\nabla \cdot \left[ \left(\mathbf{U}^H(\mathbf{r}) + Pe_f^{-1} \mathbf{U}^P(\mathbf{r})\right) g(\mathbf{r}) - \left(Pe_f^{-1} \mathbf{D}^B(\mathbf{r}) + \mathbf{D}^f(\mathbf{r})\right) \cdot \nabla g(\mathbf{r}) \right] = 0, \tag{21}$$

where $Pe_f = \dfrac{F^{ext} a}{k_b T}$. The boundary conditions are the same as the CV case, given by (13). The average mobility and interparticle force for the probe and particle 2 should be formulated in terms of $\mathbf{r}$ and $g(\mathbf{r})$. In addition $\mathbf{D}^f$ also needs to be defined. These issues are addressed immediately below in §4.3 and §4.4.

### 4.3 Average mobility and interparticle forces

The formulation of average mobility and interparticles forces is as given in our formulation for simple-shear flow which is explained in detail in earlier works [Nazockdast & Morris, 2012b, 2013]. Here we only give the final expressions. There is, however, one key difference between conditional averages in active microrheology and shear flow. In the shear flow, the motion is induced by an external straining flow, a uniform field acting on each individual particle; by contrast, in microrheology only the probe particle is subjected to an external force/velocity. As a result, when the Smoluchowski equation is integrated to pair particle level in shear flow, particles 1 and 2 are statistically identical, resulting in certain important symmetries: in shear flow $\langle \mathbf{M}_{11} \rangle_2 = \langle \mathbf{M}_{22} \rangle_2$, $\langle \mathbf{M}_{12} \rangle_2 = \langle \mathbf{M}_{21} \rangle_2$, and $\langle \mathbf{F}_2 \rangle_2 = -\langle \mathbf{F}_1 \rangle_2$. On the other hand, in the pair SE for active microrheology (whether for CF or CV), the distribution of bath particles around the probe is intrinsically different from that around particle 2 since the force or velocity is only imposed on the probe. A clear example would be the case where particle 2 and the probe are well separated. In this configuration, particle 2 is decorrelated from the probe; hence particle 2 and the bath particles around it are in the equilibrium distribution, $g(\mathbf{r_{32}}) = g^{eq}(r_{32})$ where $\mathbf{r}_{32}$ is the separation vector of particle 2 and a third bath particle. On the other hand, as it is shown in previous works and will be shown here, the distribution of bath particles around the probe is anisotropic with more particles accumulating in the direction of probe motion and a depleted zone of particles behind the probe. As a result the symmetry properties in shear flow do not hold here, and in general

$$\langle \mathbf{M}_{11} \rangle_2 \neq \langle \mathbf{M}_{22} \rangle_2, \quad \langle \mathbf{F}_2^P \rangle_2 \neq -\langle \mathbf{F}_1^P \rangle_2.$$

This effect of the lack of symmetry between the probe and bath particle of the pair becomes negligible for dilute suspensions, since the interactions are limited to the pair level (particle 1 and 2) while for concentrated suspensions it should be included in the formulation. This is taken into account through modification of the triplet distribution function, $g_3(\mathbf{r}, \mathbf{r}_{13})$, as described below.

Interparticle forces are assumed to be pair-wise additive and the conditional average forces are formulated as

$$\langle \mathbf{F}_1 \rangle_2 g(\mathbf{r}) = \mathbf{F}_{21}(\mathbf{r}) g(\mathbf{r}) + n \int \mathbf{F}_{31}(\mathbf{r}_{31}) \, g_3(\mathbf{r}, \mathbf{r}_{31}) d\mathbf{r}_{31}, \tag{22}$$

where the forces $\mathbf{F}_{21}$ and $\mathbf{F}_{31}$ are exerted on particle 1 by particles 2 and 3, respectively. The hard-sphere force is modeled here through a steep repulsion written as

$$\mathbf{F}_1^P = -(k_b T/a) \frac{\tau \exp(-\tau \epsilon)}{1 - \exp(-\tau \epsilon)} \hat{\mathbf{r}}, \tag{23}$$

where $\epsilon = r/a - 2$ is the dimensionless gap between the particle surfaces. The parameter $\tau$ determines the steepness of the repulsive force; larger $\tau$ results in a steeper decay of the repulsion, and we have used $\tau = 400$ for all the calculations here. It is necessary to express the triplet distribution, $g_3$, in terms of the pair distribution function to obtain a relationship which is only a function of $g(\mathbf{r})$ and $\mathbf{r}$. This is done here through a modified Kirkwood superposition closure approximation Rice & Lekner [1965] given by

$$g_3(\mathbf{r}, \mathbf{r}_{31}) = g(\mathbf{r}) g(\mathbf{r}_{31}) g(\mathbf{r}_{32}) \exp(\tau^\star(\mathbf{r}, \mathbf{r}_{31}, \phi)). \tag{24}$$

This modifies the Kirkwood superposition through the exponential term which is tabulated based on separation distance ($r_{31}$ here) and various values of $\phi$ [Rice & Lekner, 1965]. To take into account the fact that configurations of bath particles around particle 2 and the probe are generally different from one another this closure is modified to

$$g_3(\mathbf{r}, \mathbf{r}_{31}, \mathbf{r}_{32}) = g(\mathbf{r})g(\mathbf{r}_{31})\bigl(\lambda g^*(\mathbf{r}_{32}) + (1-\lambda)g^{eq}(r_{32})\bigr)\exp(-\tau^*), \qquad (25)$$

where $\lambda$ is a measure of connectivity of particles 2 and 3 with the probe (particle 1). This is modeled as

$$\lambda(\mathbf{r}, \mathbf{r}_{31}, \mathbf{r}_{32}) = \frac{\max\left(|\mathbf{U}_2|, |\mathbf{U}_3|\right)}{|\mathbf{U}_1|}, \qquad (26)$$

where $\mathbf{U}_1$, $\mathbf{U}_2$, and $\mathbf{U}_3$ are velocities of particles 1, 2, and 3 at each triplet configuration. This expression gives a correct description of the limits where particles are touching or well-separated. For example, if both 2 and 3 are well separated from the probe particle, $\lambda \to 0$ and the distribution of bath particles around particle 2 is the equilibrium distribution, $g^{eq}(r_{32})$. Also when all three particles are connected in the direction of the applied force $\lambda \to 1$ since three particles move as a single body along the force direction. Unlike the shear flow, $g(\mathbf{r}_{23}) \neq g(\mathbf{r}_{32})$. To correctly account for this, in (26) the separation vector of particles 2 and 3 is measured from the particle closer to 1 to the more distant particle. For example, in the case of having particle 3 between particles 1 and 2, $g^*(\mathbf{r}_{32}) = g(\mathbf{r}_{32})$; when particle 2 is between 1 and 3, $g^*(\mathbf{r}_{32}) = g(-\mathbf{r}_{32})$. In other words, between particles 2 and 3, the one closer to the probe carries the information on the motion of the probe to the particle at a greater distance. We note that this is a simple phenomenological model. Formulations of the average resistance and mobility are identical to those given in our previous paper [Nazockdast & Morris, 2013] and are not reproduced.

### 4.4 Force-induced diffusion

Variations of $\mathbf{D}_{11}^f - \mathbf{D}_{21}^f$ with separation are modeled in a similar fashion to that used in our theory for structure and rheology in shear flow [Nazockdast & Morris, 2012b, 2013]:

$$\mathbf{D}_{11}^f - \mathbf{D}_{21}^f = D^f(\mathbf{r}) = \mathbf{D}_s^f\Bigl(\mathcal{G}(r)\hat{\mathbf{r}}\hat{\mathbf{r}} + \mathcal{H}(r)\left(\mathbf{I} - \hat{\mathbf{r}}\hat{\mathbf{r}}\right)\Bigr), \qquad (27)$$

where $D_s^f$ is the isotropic portion of long-time self-diffusion of the probe, $D_s^f = tr(\mathbf{D}_s^f)/3$ which scales as $M_0 F^{ext} a$ at $Pe_f \gg 1$. The functions $\mathcal{G}(r)$ and $\mathcal{H}(r)$ determine the variations of force-induced relative diffusivity, $\mathbf{D}^f$, along the line of centers and the plane perpendicular to the pair radial vector, i.e. the local $(\hat{\theta} - \hat{\varphi})$ plane, respectively. As was done for sheared suspensions, we model the relative dispersion based on the behavior of $\mathbf{D}^f$ in the limits of a pair that is well-separated, $r/2a \gg 1$, and very near contact, $r \to 2a$. When the pair are well-separated, the particles become decorrelated and the induced dispersion on particle 2 due to motion of the probe vanishes, $\mathbf{D}_{21}^f(\mathbf{r}) = 0$, and $\mathbf{D}_{11}^f(\mathbf{r}) \to \mathbf{D}_s^f$ where $\mathbf{D}_s^f$ is the long time average self-diffusivity of the probe. The relative dispersion, here a relative diffusivity, is proportional to relative velocity fluctuations, i.e. $\mathbf{D}^f(\mathbf{r}) \propto \tilde{\tau}\langle\mathbf{U}'(\mathbf{r})\mathbf{U}'(\mathbf{r})\rangle_2$, where $\mathbf{U}'$ is the deviation of the local relative velocity from the average value and $\tau$ is the characteristic time scale over which relative motion becomes decorrelated. When the probe and particle 2 are in near-contact configurations, the relative radial velocity and its fluctuations approach zero due to the excluded volume effect, $\mathbf{U} \propto r - 2a$ and $\mathbf{U}' \propto r - 2a$. Thus the relative diffusivity in the radial direction is expected to approach zero as $(r - 2a)^2$. The excluded volume restriction is imposed along the line of centers and the fluctuations of relative relative velocity in the angular direction remain on the order of the probe velocity. As a result, at all separations we take $\mathbf{D}^f \propto M_0 F^{ext} a$ for the $\hat{\theta}$ and $\hat{\varphi}$ components.

Under equilibrium conditions, the collective diffusivity is related to osmotic pressure by $D_c^{eq} = M_c^{eq} \dfrac{\partial \Pi^{eq}}{\partial n}$ [Dhont, 1996] where subscript $c$ refers to collective diffusion and mobility. Since mobility and pressure are still well-defined outside equilibrium, one may extend the equilibrium relationship to non-equilibrium systems by analogy and write $\mathbf{D} = \mathbf{M} \cdot \dfrac{\partial \Sigma}{\partial n}$ where $\Sigma$ is the stress induced by the particle phase. Microstructural theories for predicting single particle and collective diffusion of sheared suspension also find a general correspondence between the shear induced stress and diffusion [Brady & Morris, 1997; Leshansky & Brady, 2005; Leshansky et al., 2008]. [Zia & Brady, 2012] extended this idea to active microrheology to relate probe diffusivity to the force-induced stress on the probe. They considered a phase composed of a suspension of probe particles where the suspension is so dilute that the probes have no interaction among themselves. The momentum balance is written for this phase and an exact relationship between the force-induced stress and diffusion for a dilute suspension is derived. Their analysis, however, neglects hydrodynamic interactions between the probe and the bath particles which makes a direct extension of their theory to the problem under study difficult. Instead of their detailed formulation, we use the analogy in the simplest form to compute the limiting value of $\mathbf{D}^f(\mathbf{r})$ when the probe and particle 2 are well-separated, $\mathbf{D}_s^f$ and write

$$\mathbf{D}_c^f = \mathbf{D}_s^f = \mathbf{M} \cdot \frac{\partial \Sigma}{\partial n_a}, \tag{28}$$

where $n_a$ is the number density of the probe phase ( $n_a \ll 1$) and $\Sigma$ is the average stress on the probe phase per unit volume. Note that since the probe phase is considered very dilute (no interactions between the probes), the collective diffusivity and mobility become identical to their single particle values in (28).

In the next step, we take the enhanced diffusion to be dominated by hydrodynamic velocity fluctuations (at $Pe_f \gg 1$ where the effect is pronounced) and hence only consider the hydrodynamic contribution to the stress in (28). The reasonable assumption that the hydrodynamic stress is dominated by near-contact lubrication stress allows $\Sigma$ to be written in terms of the moment of external force in the radial direction:

$$\Sigma^H = n_a n \int_{2a}^{r^*} \left( \mathbf{F}^{ext} \cdot \mathbf{r} \right) \hat{\mathbf{r}} \hat{\mathbf{r}} g(\mathbf{r}) d\mathbf{r}. \tag{29}$$

The integration of stress is carried out inside the boundary layer of $g(\mathbf{r})$ from contact to $r = r^*$, where $r^*$ is taken to be the radius at which the increase in angularly averaged pair correlation function, $\tilde{g}(r)$, from the equilibrium distribution drops to 10% of the contact values i.e. $\tilde{g}(r^*) = 0.1\left[\tilde{g}(2a) - g^{eq}(2a)\right] + g^{eq}(r^*)$. This is similar to the definition used by Sierou & Brady [2002] as well as our earlier work on shear flow [Nazockdast & Morris, 2013]. Note that $\mathbf{D}^f$ has no dependence on $n_a$ after taking the derivative with respect to $n_a$ in (28). From (28) and (29) the force-induced diffusivity is found to scale roughly as $D_s^f \sim U^{ext}\phi a g(2)$ where $g(2)$ is the contact value of pair distribution function in the boundary layer portion of the contact surface.

For simplicity we assume the variations of relative diffusion with $\phi$ take the same form as the far-field value, $D_s^f$, for all pair separations. As a result $\mathcal{G}$ and $\mathcal{H}$ are only functions of $r$. The near-contact relative radial diffusion caused by the force on the probe varies as $(r-2)^2$ (with lengths scaled by $a$) near contact and approaches the force-induced self diffusion when the probe and particle 2 are well separated. Based on these limits $\mathcal{G}$ is modeled as

$$\mathcal{G}(r) = 1 - \exp\left[-\tilde{m}(r/a - 2)^2\right], \tag{30}$$

where $\tilde{m}$ is a constant for all conditions of $\phi$ and $Pe_f$: this is the only fitting parameter in our theory. Setting $\tilde{m} = 0.01$ resulted in a good agreement between theory and simulation. We did not attempt to systematically vary to find the optimal value of $\tilde{m}$.

Since there is little experimental or simulation data on active microrheology to provide meaningful averages of the probe and bath dynamics, we use the insight from shear flow Accelerated Stokesian Dynamics simulations for modeling the variations of angular diffusivity. The results of simulation sampling at $Pe \gg 1$ show the magnitude of shear-induced angular relative velocity fluctuations remains finite and is reduced to approximately 0.25 of the maximum far-field value [Nazockdast & Morris, 2013]. We assume these limits hold and model $\mathcal{H}$ as

$$\mathcal{H}(r) = \frac{3}{4}\left(1 - \exp\left[-\tilde{m}(r-2)^4\right]\right) + \frac{1}{4}. \tag{31}$$

The conditional averages appearing in (11) and (19) are expressed in terms of integrals of $g(\mathbf{r})$. The resulting pair Smoluchowski equations for both CF and CV conditions are nonlinear integro-differential equations which we solve using a finite element iterative scheme with the boundary conditions given in (13). The details of the solution technique are given in Nazockdast & Morris [2012b]. There is, however, one difference in computation domain of the active microrheology and shear flow. In shear flow, the structure changes in all three directions in space while active microrheology is an axisymmetric problem with symmetry about the direction of external force on, or velocity of, the probe.

# 5 Viscosity

In this section, we develop the relationship between $g(\mathbf{r})$ and apparent viscosity which is calculated using the Stokes-Einstein relationship, $\eta_{micro} = \dfrac{F}{6\pi U a}$. In a CF experiment, $F = F^{ext}$ is known and the goal is to compute $U^{ext}$ as a function of $g(\mathbf{r})$, implying the need for the average probe mobility, $\langle \mathbf{M}_{11} \rangle_2$. For CV, the problem is inverted: the velocity is fixed and the external force should be calculated which requires computation of the average resistance to motion of the probe.

## 5.1 Constant velocity viscosity, $\eta_U$

The motion of the probe at any separation from particle 2 due to Brownian motion is the sum of the forces directly applied from the fluid, $\langle \mathbf{F}_1^B \rangle_2 = k_b T \nabla g / g$ and the hydrodynamic interaction owing to the Brownian motion of particle 2 given by $-\mathbf{R}_{12} \cdot \mathbf{U}_2^B$ where $\mathbf{U}_2^B = -k_b T (\mathbf{R}_{22})^{-1} \cdot \nabla g / g$ is the Brownian velocity of particle 2. Combining these relationships, the total Brownian force on the probe simplifies to

$$F_z^B = n\, k_b T \int \left[ \mathbf{I} + \mathbf{R}_{12} \cdot (\mathbf{R}_{22})^{-1} \right] \cdot (\nabla g / g) \cdot \hat{\mathbf{z}} g(\mathbf{r}) d\mathbf{r} = n\, k_b T \int (\nabla \cdot \mathcal{C}) \cdot \hat{\mathbf{z}} g(\mathbf{r}) d\mathbf{r, \tag{32a}}$$

where subscript $z$ implies the component of the force along the direction of probe motion and $\mathcal{C} = -\mathbf{R}_{22}^{-1} \cdot \mathbf{R}_{21} - \mathbf{I}$. Integration by parts and the divergence theorem were used to recast the second integral form from the first expression on the right hand side for $F_z^B$ in (32a). A similar formulation is used to obtain the net interparticle force on the probe resulting in

$$F_z^P = -n\, k_b T \int \left[ \mathcal{C} \cdot \mathbf{F}_{21}^P(\mathbf{r}) \right] \cdot \hat{\mathbf{z}} g(\mathbf{r}) d\mathbf{r}. \tag{32b}$$

Finally the hydrodynamic force applied from the fluid on the probe is simply the product of the average resistance of the probe in the direction of motion and its velocity:

$$F_z^H = -\langle R_{11}^{zz} \rangle U^{ext}. \tag{32c}$$

The summation of the forces in (32) gives the force applied on the probe from the fluid and bath particles. Thus $F_U^{ext} = -\left(F_z^B + F_z^P + F_z^H\right)$ needs to be externally applied to the probe to maintain its constant velocity of $U^{ext}$. The integrals in (32) are easily computed once $g(\mathbf{r})$ is known from the theory. The micro-viscosity is calculated as

$$\frac{\eta_U}{\eta} = \hat{\eta}_U = \frac{-\left(F_z^B + F_z^P + F_z^H\right)}{6\pi \eta U^{ext} a}. \tag{33}$$

We only consider hard-sphere interparticle interactions. Thus, $\mathbf{F}^P$ is only non-zero at contact. The hard-sphere force is modeled using a steep repulsive radial force, (23). Using simple scaling arguments, it is straightforward to show that $F_z^P \propto F_z^B \tau^{-1}$. Hence the interparticle force is neglected here in comparison with Brownian forces. This assumption does not generally hold for particles interacting through long-ranged soft potentials and the contribution from interparticle interaction needs to be computed Nazockdast & Morris [2012a]. The hydrodynamic dimensionless viscosity for hard-sphere interaction is thus

$$\hat{\eta}_U^H = \langle R_{11}^{zz} \rangle / R_0, \tag{34}$$

where $R_0 = 6\pi\eta a$ is the single particle resistance. The Brownian micro-viscosity is $\hat{\eta}_U^B = \hat{\eta}_U - \hat{\eta}_U^H$.

## 5.2 Constant force viscosity, $\eta_f$

The CF micro-viscosity is defined as

$$\eta_f = \frac{F^{ext}}{6\pi \langle U^{ext} \rangle a}, \tag{35}$$

where $\langle U^{ext} \rangle$ is the average velocity of the probe particle in the direction of the applied force. The average probe velocity is stated in terms of contributions from hydrodynamic, interparticle and Brownian interactions, i.e.

$$\langle \mathbf{U}^{ext} \rangle = \mathbf{U}^H + \mathbf{U}^P + \mathbf{U}^B, \tag{36}$$

with the contributions related to $g(\mathbf{r})$ through the following relations [Khair & Brady, 2006]:

$$\mathbf{U}^H = \langle \mathbf{M}_{11} \rangle \cdot \mathbf{F}_1^{ext}, \tag{37a}$$

$$\mathbf{U}^P = n \int \langle \mathbf{M}_{11} - \mathbf{M}_{12} \rangle_2 \cdot \mathbf{F}_{21}^P(\mathbf{r}) g(\mathbf{r}) d\mathbf{r}, \tag{37b}$$

$$\mathbf{U}^B = -n \int \nabla \cdot \langle \mathbf{M}_{11} - \mathbf{M}_{12} \rangle_2 g(\mathbf{r}) \, d\mathbf{r}. \tag{37c}$$

The coefficients appearing in (37a - 37c) are known through solving for $g(\mathbf{r})$. For near hard-sphere dispersions, the contribution of bath particles to viscosity is taken to be a summation of contributions from pure fluid, Brownian, and hydrodynamic interactions. The hydrodynamic contribution to viscosity is defined as

$$\frac{\eta_f^H}{\eta} = \hat{\eta}_f^H = \frac{M_0}{\langle M_{11}^{zz} \rangle}, \tag{38}$$

where $\langle M_{11}^{zz} \rangle$ is the average probe mobility in the force direction. The Brownian viscosity is obtained by subtracting hydrodynamic viscosity from the total viscosity: $\hat{\eta}^B = \hat{\eta}_f - \hat{\eta}_f^H$.

# 6 Predictions of the theory

In presenting the structure at weak $Pe$, or near-equilibrium conditions, the theoretical results presented in §6.1 stand alone because statistically meaningful results were not accessible in a reasonable time by simulation. The results of theory and simulation at $Pe \geq 1$ are compared in §6.2.

## 6.1 Near-equilibrium, $Pe \ll 1$

Near equilibrium, the distortion of microstructure from its equilibrium form can be formally expressed as a regular perturbation expansion in terms of $Pe$ as

$$g^U(\mathbf{r}) = g_{eq}(r) + Pe_U g_1^U(\mathbf{r}) + \mathcal{O}(Pe_U^2), \tag{39a}$$

$$g^f(\mathbf{r}) = g_{eq}(r) + Pe_f g_1^f(\mathbf{r}) + \mathcal{O}(Pe_f^2), \tag{39b}$$

with

$$g_1^U(\mathbf{r}) = f_1^U(r) U^{ext} \hat{\mathbf{z}} \cdot \hat{\mathbf{r}} \tag{39c}$$

$$g_1^f(\mathbf{r}) = f_1^f(r) F^{ext} \hat{\mathbf{z}} \cdot \hat{\mathbf{r}}, \tag{39d}$$

where subscripts $U$ and $f$ present the constant velocity and force conditions. When $Pe \to 0$, $\mathcal{O}(Pe^2)$ contributions to the structure become negligible, and the zero-force/velocity viscosity is determined by $f_U^1$ and $f_f^1$. To calculate $f_1^f$ and $f_1^U$, we compute $g(\mathbf{r})$ for $\phi \leq 0.40$ at $Pe_f$ and $Pe_U = 0.01$, for which contributions of $\mathcal{O}(Pe^2)$ and higher order terms can be neglected. Note that $\phi = 0.40$ is the largest volume fraction for which the modified Kirkwood superposition gives a convergent solution for $g(\mathbf{r})$ [Rice & Lekner, 1965]. The expressions (39c) and (39d) are then fitted with good agreement to the predicted microstructures at $Pe_{f,U} = 0.01$ to give $f_1^U(r)$ and $f_1^f(r)$. To study the effect of CF and CV conditions, the structures should only be compared against each other when they both correspond to the same external force or velocity. The choice of $\mathbf{F}^{ext} = \langle \mathbf{R}_{11} \rangle \cdot \mathbf{U}^{ext}$, results in $\langle U^{ext} \rangle = U^{ext}$. This constraint is satisfied by rescaling the Péclet number for CF as $Pe_{\langle U \rangle} = Pe_f (\hat{M}_{11}^{zz})^{-1}$ which is analogous to comparing $f_1^U(r)$ against $f_1^f(r)(\hat{M}_{zz}^{11})^{-1}$ where $\hat{M}_{11}^{zz} = M_{11}^{zz}/M_0$.

Figure 5(a) shows the variations of $f_U^1(2)$ and $f_f^1(2)(\hat{M}_{11}^{zz})^{-1}$ with $\phi$, showing that the values increase with volume fraction and the distortions are slightly stronger in CV. The difference is stronger at higher $\phi$. Having roughly similar structural distortion in the CF and CV scenarios is not in agreement with the predictions of Squires & Brady [2005]for dilute Brownian suspensions with no hydrodynamic interactions (HI), in which a distortion twice as large for constant velocity compared to constant force was predicted. Recall that both probe and the bath particles diffuse due to Brownian fluctuations in CF while in CV only the bath particle does so. This gives rise to a factor of two enhancement to the pair diffusion in CF compared with CV when HI are ignored causing the distortion of pair microstructure in CV to be twice as large as CF conditions. At finite volume fractions, the presence of many-body interactions change this simple analysis. In the equilibrium limit, these interactions introduce a potential of mean force, whose effects include the secondary peaks in pair correlation function, and nonlinear variations of material properties such as the viscosity, the osmotic pressure, and the diffusion coefficient with volume fraction. Below we give a brief explanation of these effects and demonstrate how they alter the structure at $Pe \ll 1$ with respect to the predictions of dilute theories.

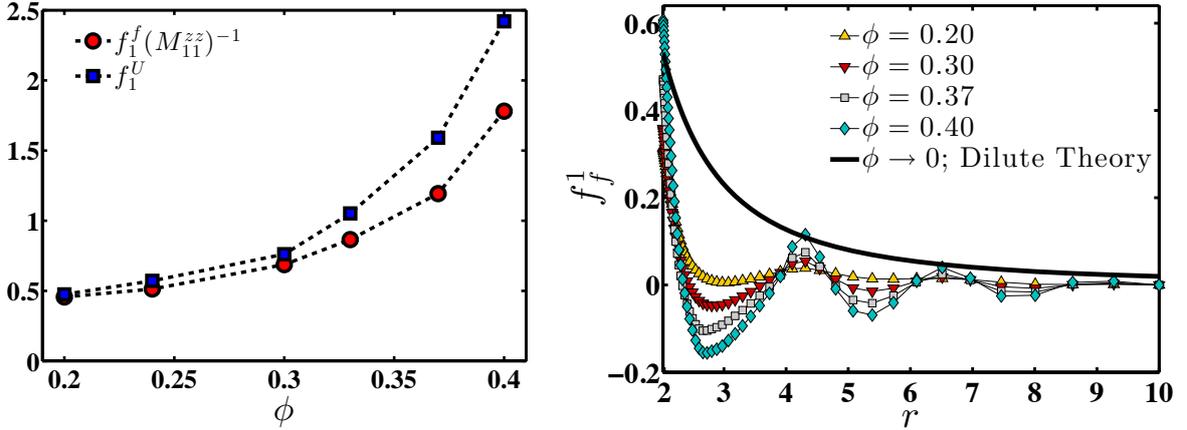

**Figure 5:** *(a) The contact values of $\mathcal{O}(Pe)$ structural deformation functions for CF and CV, $(\hat{M}_{11})^{-1}f_f^1(2)$ and $f_U^1(2)$ in (39a,b), vs $\phi$; (b) the variations of $f_f^1$ with radial separation, $r$, at different volume fractions compared against the predictions of the dilute theory (equation (4.9) in Khair & Brady [2006]).*

In figure 5(b) we compare the radial variations of $f_1^f$ at different volume fractions against the predictions of the dilute theory (equation (4.9) in Khair & Brady [2006]). The main observation is that the decay of structural distortion away from contact is significantly faster in the predictions of the full theory than the dilute theory, even at the lowest volume fraction studied, $\phi = 0.20$. The structural buildup at contact from the present concentrated theory drops rapidly and oscillates about zero for $r > 2.5$, whereas in the dilute theory the distortion is monotonically decreasing and is still $1/4$ of the value at contact when $r = 4$. The secondary peaks in $f_1^f$ are generated by propagation of the anisotropic pair correlation buildup near contact through interactions with the bath particles [Nazockdast & Morris, 2012b]; a similar propagation of correlation in equilibrium results in the secondary peaks in $g_{eq}(r)$. The sharp decay of structural anisotropy compared to the dilute limit, and the presence of secondary peaks in $f_1^f$ both show that the near contact interactions determine the structure at volume fractions as low as $\phi = 0.20$.

The extent of distortion of equilibrium structure at small $Pe$ is proportional to the ratio of convective to diffusive fluxes. In figure 6(a), we compare the dimensionless relative diffusion coefficients from CF, $2(\mathbf{M}_{11}-\mathbf{M}_{21})$, and CV, $\mathbf{M} = \mathbf{R}_{22}^{-1}$, at $\phi = 0.33$. At large separations, as expected, the relative diffusion of CF conditions is twice as large as the CV case while in both cases, the values asymptote to zero at contact. In figure 6(b) the variations near contact, in $2 \leq r < 2.5$, are considered. We have also included the variations of relative diffusion coefficients of an isolated pair in the dilute theories for comparison. For both the dilute and concentrated suspensions in this range of pair separations, the ratio of CF to CV relative diffusivities is at most 1.2 compared to 2 in Brownian suspensions with no HI. This trend was observed for other volume fractions as well. Since the structural distortion predominantly occurs in $r < 2.5$ as shown by figure 5(b), where the relative diffusion coefficients for CF and CV are quantitatively close, our theory predicts similar distortions for the two conditions at a given average velocity; see figure 5(a).

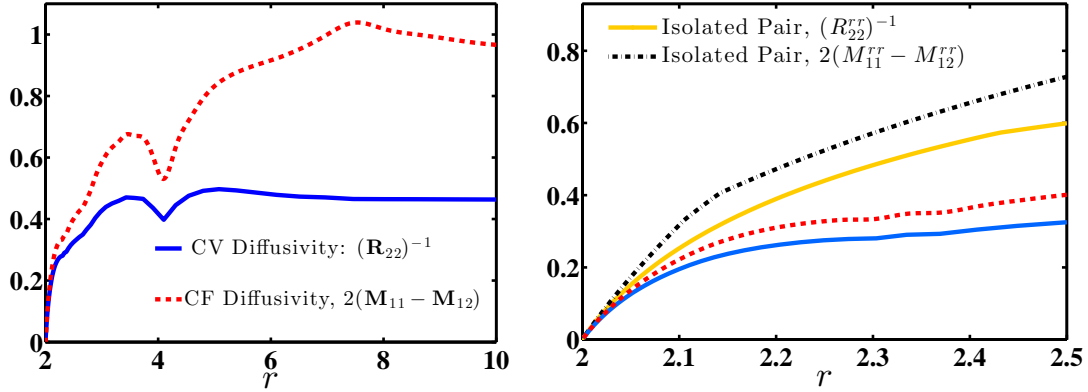

**Figure 6:** *(a) The variations of the dimensionless relative diffusivities with pair separation distance, r, under CF and CV conditions and $\phi = 0.33$. At $r \gg 1$, the CF diffusion is twice the value of the CV diffusion while both diffusivities approach zero at contact; (b) the results of figure 6(a) shown for $2 \leq r \leq 2.5$. The isolated pair relative diffusion coefficients from the dilute theory are also shown for comparison.*

For further insight, we analyze the governing equations for $g_U^1$ and $g_f^1$. The equations, obtained by substituting (39a) and (39b) in (11) and (21), are

$$\nabla \cdot \left[ (\mathcal{C} \cdot \hat{\mathbf{z}}) g_{eq}^U + \mathbf{U}_{eq}^U g_1^U + \mathbf{U}_1^U g_{eq}^U - \mathbf{M} \cdot \nabla g_1^U \right] = 0, \tag{40a}$$

$$\nabla \cdot \left[ (-\mathbf{M}_r \cdot \hat{\mathbf{z}}) g_{eq}^f + \mathbf{U}_{eq}^f g_1^f + \mathbf{U}_1^f g_{eq}^f - 2\mathbf{M}_r \cdot \nabla g_1^f \right] = 0, \tag{40b}$$

where $\mathbf{M}_r = \mathbf{M}_{11} - \mathbf{M}_{12}$. Similar to perturbation expansion of $g(\mathbf{r})$, we have expanded the velocity induced by interparticle interactions, $\mathbf{U}$, as

$$\mathbf{U}^{f,U} = \mathbf{U}_{eq}^{f,U} + Pe\,\mathbf{U}_1^{f,U}, \tag{40c}$$

with

$$\mathbf{U}_{1,eq}^U = \mathbf{M} \cdot (\mathbf{F}_2^P)_{1,eq}, \tag{40d}$$

$$\mathbf{U}_{1,eq}^f = \mathbf{M}_r \cdot (\mathbf{F}_2^P - \mathbf{F}_1^P)_{1,eq}, \tag{40e}$$

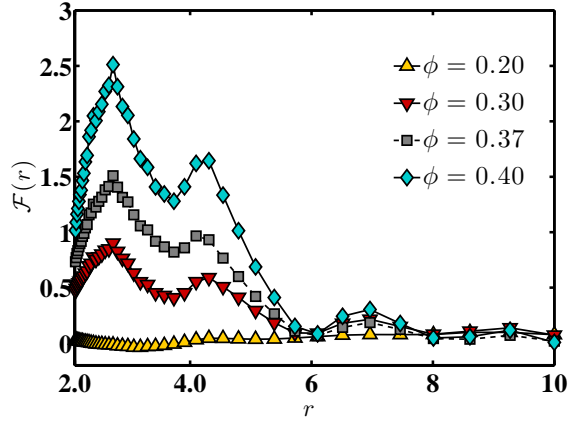

**Figure 7:** *The ratio of $\mathcal{O}(Pe)$ convective fluxes from interparticle interactions, $(\mathbf{U}^P)_1^U g_{eq}$, to the flux induced by the average motion of the probe, $(\mathcal{C} \cdot \hat{\mathbf{z}}) g_{eq}$, under CV conditions as a function of radial separation; see equation 41.*

where $\mathcal{C} = -(\mathbf{R}_{22})^{-1} \cdot \mathbf{R}_{21} - \mathbf{I}$. For readability we have dropped $\langle . \rangle_2$ around conditionally averaged quantities. The solution to the pair Smoluchowski equation in equilibrium can also be used to write $(\mathbf{F}_2^P)_{eq} = -(\mathbf{F}_1^P)_{eq} = -\nabla \log g_{eq}(r)$ which implies this is the force arising from the well-known potential of mean force in liquid-state theory; the many-body flux terms in (40a) and (40b) are not present in the dilute theories. One measure of the importance of these particle interactions can be obtained by taking the ratio of the $\mathcal{O}(Pe)$ anisotropic convection flux induced by interparticle interactions, $(\mathbf{U}_{eq}^U) g_1^U + (\mathbf{U}_1^U) g_{eq}^U$, to the convection flux induced by average motion of the probe, $(\mathcal{C} \cdot \hat{\mathbf{z}}) g_{eq}$. The predictions of the theory show that generally $|\mathbf{U}_{eq}^U g_1^U| \ll |\mathbf{U}_1^U g_{eq}^U|$. In this case the ratio of two convection fluxes in radial direction simplifies to

$$\Gamma = \left(\mathbf{U}_1^U \cdot \hat{\mathbf{r}}\right) (\hat{\mathbf{r}} \cdot \mathcal{C} \cdot \hat{\mathbf{z}})^{-1}. \tag{41a}$$

Due to the linearity of the structural distortion with $U^{ext}$, the radial velocity is decomposable to the general form of

$$\Gamma = \hat{\mathbf{z}} \cdot \hat{\mathbf{r}} \mathcal{F}(r). \tag{41b}$$

We plot $\mathcal{F}(r)$ for various volume fractions in figure 7. At all $\phi$, the ratio is significant and for $\phi > 0.30$ it becomes larger than 1 for regions near the boundary; this is indicative of the dominating role of the underlying structure and its effect on particle interactions in setting the microstructure in the near-equilibrium dense suspension. The results for CF conditions are quantitatively close to those of figure 7.

## 6.2 Far from equilibrium, $Pe \gg 1$

For $Pe \geq 1$, simulation results are available for comparison against the theoretical predictions of microstructure. Figure 8 compares the predictions of $g(\mathbf{r})$ in the plane of symmetry with the results obtained from sampling configurations from ASD simulations at $Pe_f = 10, 100$, and 1000. At $Pe_f = 10$ a clear distortion of equilibrium structure is observed. The distortions become stronger with $Pe_f$, leading to a boundary layer structure on the leading half of the probe surface. In both theory and simulations the structure away from contact saturates at $Pe_f = 1000$ as a result of force-induced dispersion being the predominate far-field relative diffusivity. The predictions for $Pe_f = 400$ and $Pe_f = 1000$ are quite close except for the maximum values of $g(\mathbf{r})$ at contact. We do not present the $pe_f = 400$ condition in figure 8. At all Péclet numbers, the second nearest neighbor peaks are distorted similarly to $g$ at contact, i.e. the values are largest in the flow direction and smallest behind the probe. There is a general agreement between the predictions and simulation results. However, differences can be noted, including the fact that the form of the theory and simulational distortion at $Pe_f = 10$ behind the probe differ. The next nearest neighbor peaks are stronger in the predicted microstructure, indicating that the integral formulation for many-body interactions overpredicts the correlation between the particles. This was also observed in our previous study of pair dynamics in sheared suspensions [Nazockdast & Morris, 2013]. Also, in the ASD results the width of the farfield wake zone narrows with increase of pair separation distance more than is seen in the theoretical predictions.

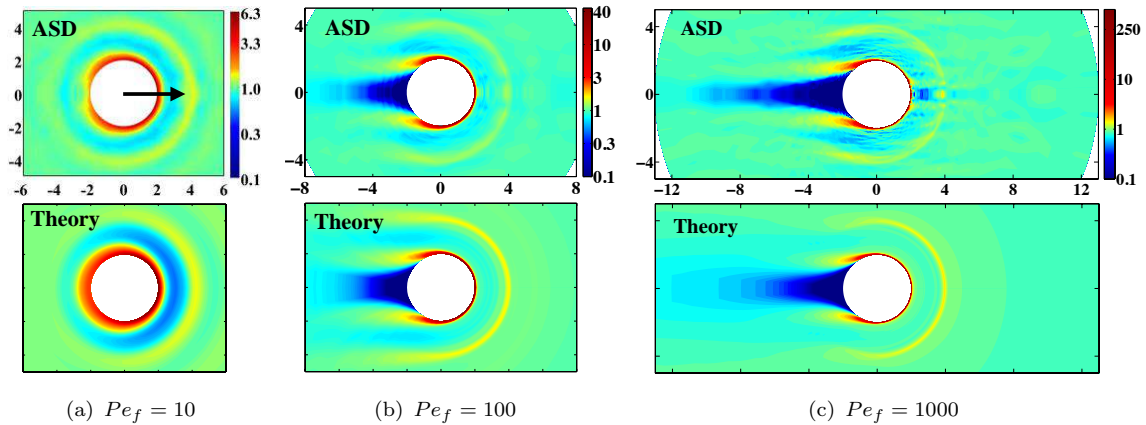

(a) $Pe_f = 10$  (b) $Pe_f = 100$  (c) $Pe_f = 1000$

**Figure 8:** *The variations of $g(\mathbf{r})$ in the $(r - z)$ plane passing through the probe particle diameter, at fixed force from simulation (ASD) and theory, for $\phi = 0.40$. (a) $Pe_f = 10$, (b) $Pe_f = 100$, and (c) $Pe_f = 1000$. Identical color bars are chosen for theory and simulation.*

To investigate the performance of the theory at different volume fractions, we compare the predictions with the sampled structures from ASD for $Pe_f = 100$ at $\phi = 0.30$ and $0.50$ in figure 9. The theory correctly predicts the reduction of the length of the wake zone with increasing $\phi$ due to the enhancement of force-induced dispersion at higher particle loading. In the theory, the length of the wake zone is determined by the ratio of convective to diffusive fluxes outside of the boundary layer. When force-induced diffusion is accounted for, the diffusion flux at $Pe_f \gg 1$ is determined by $\mathbf{D}^f \propto \mathbf{M} F^{ext} g(2)\phi$ (see §4.4). In this limit $\mathbf{j}_D/\mathbf{j}_C \propto \phi$ leading to a smaller wake at higher $\phi$, and saturation of the wake length for $Pe_f \gg 1$.

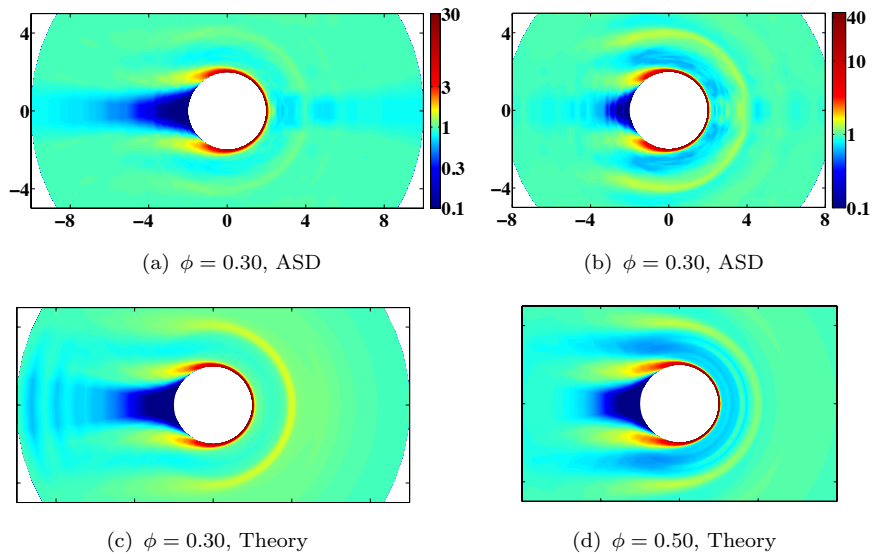

**Figure 9:** *Theory predictions and results of simulation (ASD) for $g(\mathbf{r})$ under CF conditions at $Pe_f = 100$: (a) ASD and $\phi = 0.30$, (b) theory and $\phi = 0.30$, (c) ASD and $\phi = 0.50$, and (d) theory and $\phi = 0.50$.*

Figure 10 shows $g(\mathbf{r})$ for $\phi = 0.40$ and $Pe_U = 10, 25$ and 100, from both theory and simulation. The most prominent feature seen in CV forcing, unlike CF, is that the length of the wake continuously increases with $Pe_U$, and eventually the wake spans the entire simulation box. The length of the wake at $Pe_U = 25$ is larger than $Pe_f = 1000$ which corresponds to approximately $Pe_{\langle U \rangle} = Pe_f(\hat{M}_{11}^{zz})^{-1} = 285$. Recall that this rescaling of Péclet number corresponds to equal velocities in both CF and CV conditions. The theory developed here gives reasonable predictions of structure in both CF and CV scenarios and shows that the observed differences can be attributed to the dispersive motion of the probe for CF conditions, a flux mechanism absent under CV conditions. A difference between theory and ASD results can also be noted by comparing figure 10(a) and figure 10(d). Similar to CF conditions, the theory overestimates the magnitude of the next nearest neighbor peaks.

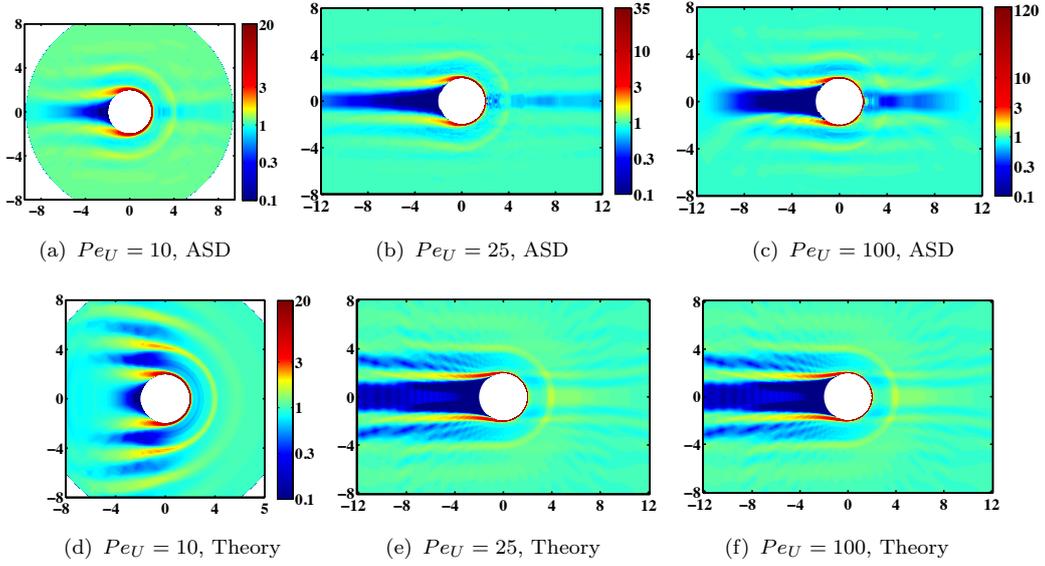

**Figure 10:** *The prediction and simulation results of $g(\mathbf{r})$ under CV conditions at $\phi = 0.40$ and varying $Pe_U$. (a) ASD and $Pe_U = 10$, (b) ASD and $Pe_U = 25$, (c) ASD and $Pe_U = 100$, (d), theory and $Pe_U = 10$, (e) theory and $Pe_U = 25$, and (f) theory and $Pe_U = 100$.*

### 6.2.1 Microstructure near the probe

The influence of probe motion on average resistance (or viscosity) is controlled by near-contact interactions of the probe with bath particles, which is quantified by the near-contact form of $g(\mathbf{r})$. The values of pair microstructure at contact, $g(2;\varphi)$, from theory and simulation are compared in figure 11 at $Pe_f = 1, 10$ and $100$ and $\phi = 0.40$. The theory gives a good quantitative prediction of simulation results in the range of $Pe$ studied.

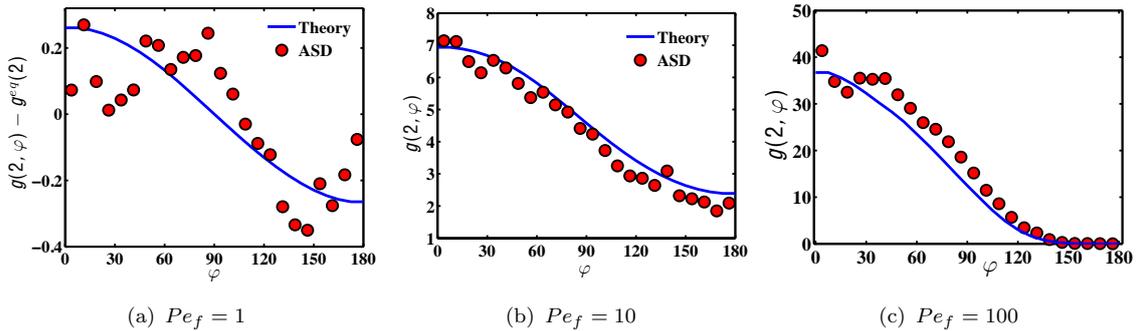

**Figure 11:** *The predictions and simulation results of pair distribution function at contact as a function of azimuthal angel, $g(2,\varphi)$, (a) $Pe_f = 1$, (b) $Pe_f = 10$, and (c) $Pe_f = 100$.*

Next we examine the effect of volume fraction. Figure 12(a) shows $g(2; \varphi)$, for $Pe_f = 100$ and $\phi = 0.30, 0.40, 0.45$ and $0.50$ from simulational sampling and predictions of the theory. The variations are almost independent of volume fraction, except for a slight deviation for $\phi = 0.30$. This can be understood by analyzing the ratio of convection to diffusion terms in pair SE. The diffusion at contact is controlled by Brownian diffusion $\mathbf{j}^D = -k_b \langle \mathbf{M} \rangle_2 \nabla g$. Since mobility appears as a multiplying factor in both the convective and diffusive flux terms, changing the volume fraction does not change the ratio of the two fluxes, $\mathbf{j}_D/\mathbf{j}_C = Pe_f^{-1} \nabla \ln g(\mathbf{r})$, and near contact structure remains almost unchanged while varying $\phi$ at a given $F^{ext}$.

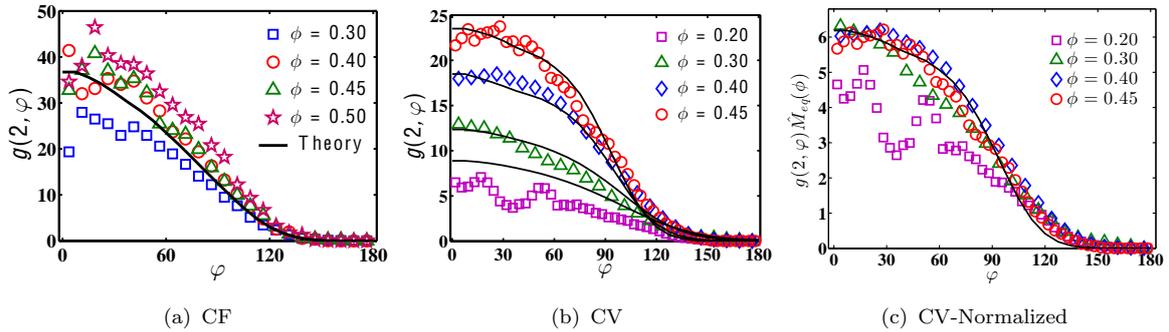

(a) CF  (b) CV  (c) CV-Normalized

**Figure 12:** *Simulation results and theory prediction for $g(2, \varphi)$ at different volume fractions for CF conditions at $Pe_f = 100$ and CV conditions at $Pe_U = 10$. (a) CF, (b) CV, (c) the results of figure 12(b), multiplied by the dimensionless equilibrium mobility of the probe at the given volume fraction, $\hat{M}_{eq}(\phi) = M_{eq}(\phi)/M_0$. The solid lines are the predictions of the theory. In figures 12(a) and 12(c), only the predictions at $\phi = 0.40$ are illustrated. Predictions in other volume fractions are quantitatively very similar and are not presented to aid visualization.*

In figure 12(b) we illustrate $g(2, \varphi)$ at $\phi = 0.20, 0.30, 0.40$ and $0.45$ for CV conditions. To avoid interaction with the periodic image in ASD simulation, $Pe_U = 10$ was chosen for all volume fractions. The predictions are presented with solid lines. There is good agreement between theory and simulation results. Figure 12(c) presents the same results with the pair distribution function multiplied by equilibrium dimensionless mobility. The curves collapse well on a single curve, with some deviation at $\phi = 0.20$. Similar to CF conditions, the near contact pair microstructure is controlled by the force required to pull the probe. Note that the force-induced diffusion becomes smaller than Brownian diffusion inside the boundary layer; hence we expect similar scaling of near contact microstructure with the applied force (or average velocity) and with $\phi$ for both CF and CV conditions.

Figure 13 shows the maximum contact value of the pair distribution function as a function of $Pe_U$ for CV and the equivalent Péclet number for CF based on the average probe motion, $\langle U^{ext} \rangle$. The equilibrium contact value of $g$ was subtracted to consider only the structural distortion due to external forcing. When CF values are plotted against $Pe_{\langle U \rangle}/2$, they lie on the same line as CV values, i.e. the near contact microstructures of CV and CF conditions are matched when they are compared at equal $Pe_U$ and $Pe_{\langle U \rangle}/2$. The factor of $1/2$ was multiplied to $Pe_{\langle U \rangle}$ based on the arguments given by the previous dilute theories stating that in CV only the bath particles can move by Brownian motion, making the relative diffusivity half that of the CF conditions. Although the details of hydrodynamic interactions may change the shifting factor by some amount, the analysis of dilute suspensions seems to give a good representation of microstructure at contact. Thus we find that the main shortcoming of the dilute theories at $Pe \gg 1$ is the absence of force-induced diffusion which results in incorrect predictions of pair microstructure away from contact under CF conditions.

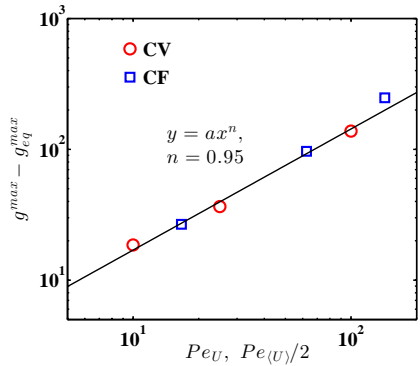

**Figure 13:** *Simulation results for increase in maximum contact value of pair distribution function from equilibrium, $g^{max} - g^{max}_{eq}$, as a function of $Pe_U$ for CV and $Pe_{\langle U \rangle}/2$ for CF. The solid line is the best power-law fit to the data with exponent of $n = 0.95$.*

## 7 Microrheology

Figure 14 shows the predicted dimensionless hydrodynamic, Brownian, and total micro-viscosity as a function of $\phi$ for CF and CV. The Accelerated Stokesian Dynamics simulation values (obtained using a Green-Kubo relation) are also presented for comparison. The predicted values for zero-force/velocity hydrodynamic micro-viscosity are identical for both CF and CV since the average mobility near equilibrium is independent of $Pe$ and is defined by $g^{eq}(r)$ (recall that the hydrodynamic viscosity is the inverse of the average probe mobility). Hence only CV predictions are presented.

It is evident that the values of Brownian viscosities are significantly smaller than shear flow in all volume fractions. We note that the Brownian viscosities for CF and CV are quantitatively close, with CV values becoming slightly larger with increasing $\phi$. This is similar to the trend observed for $\mathcal{O}(Pe)$ structural deformations in figure 5(a). Since under both conditions $\eta^B$ is determined as an integral of the product of structural anisotropy and divergence of the pair velocity in pair space (see (32a) and (37c)), similar structural distortions result in similar Brownian viscosities. We again observe a contrast between our predictions and the results of dilute theory theories that predict $\eta_U^B \approx 2\eta_f^B$ [Squires & Brady, 2005; Swan & Zia, 2013].

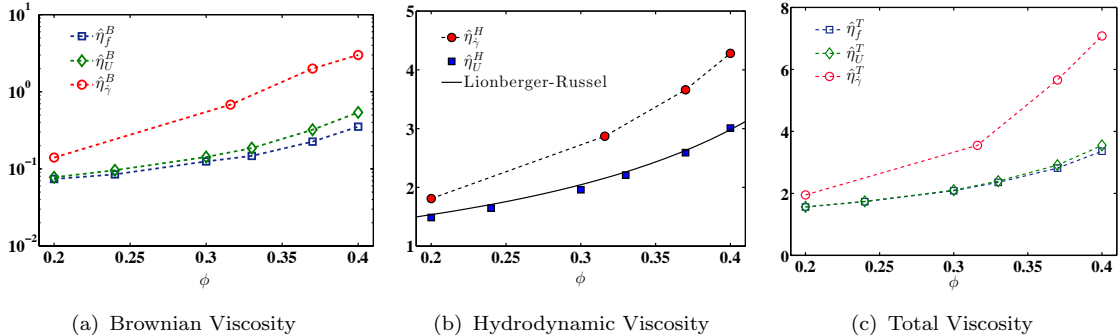

(a) Brownian Viscosity

(b) Hydrodynamic Viscosity

(c) Total Viscosity

**Figure 14:** *The predicted near-equilibrium CF and CV viscosities as a function of $\phi$ compared against Accelerated Stokesian Dynamics (ASD) simulation results for shear flow, (a) the Brownian viscosity, (b) the hydrodynamic viscosity, and (c) the total viscosity given by the summation of the Brownian and hydrodynamic viscosities.*

Figure (14(b)) shows that the predicted near-equilibrium hydrodynamic viscosity is smaller than the values obtained in shear flow for all volume fractions. There is a very good agreement between $\hat{\eta}^H$ and the solid line in Figure 14(b), which presents the empirical relationship for the inverse of the dimensionless short-time self-diffusion coefficient at equilibrium, $D_0/D_s^s(\phi) = [(1 - 1.56\phi)(1 - 0.27\phi)]^{-1}$. This result was given by Lionberger & Russel [1997] based on fitting to experimental measurements of Pearson & Shikata [1994] on hard-sphere suspensions near equilibrium, with $D_0 = k_b T M_0$ the isolated particle short-time self-diffusion; recall that $\hat{\eta}^H = M_0/\langle M \rangle = D_0/D_s^s(\phi, Pe)$. This indicates that the zero force/velocity mobility of the probe is predicted accurately for $\phi \leq 0.40$.

Figure 15 shows variations of theoretical predictions of CF and CV viscosities with the relevant Péclet number for a $\phi = 0.40$ hard-sphere suspension along with our active microrheology ASD simulation results for total viscosity. The results are compared against the simulation results of viscosity in shear flow at different $Pe_{\dot{\gamma}} = \frac{6\pi\eta\dot{\gamma}a^3}{k_bT}$. The ASD results for microrheology are only presented for a limited range of $Pe$ due to the difficulty of getting statistically reliable data at low $Pe$ and the large simulation box size needed for $Pe \gg 1$, particularly for CV conditions. For these limited data points, ASD results are in good agreement with predictions for viscosity. The hydrodynamic contributions to viscosity under both CF and CV predictions follow the trend observed in the shear flow results and show very little change with Péclet number until $Pe = 1$. At larger Péclet numbers, the values mildly increase with $Pe$ in the range studied here. The predicted values for $\hat{\eta}^H$ are very close for CF and CV, reflecting the similar near-contact microstructure. Brownian contributions in all cases are reduced with increase of $Pe$ and become negligible at $Pe > 1$. The total viscosity for both CF and CV conditions shows very little force thickening, even less than the mild shear thickening seen in the simulated shear flow.

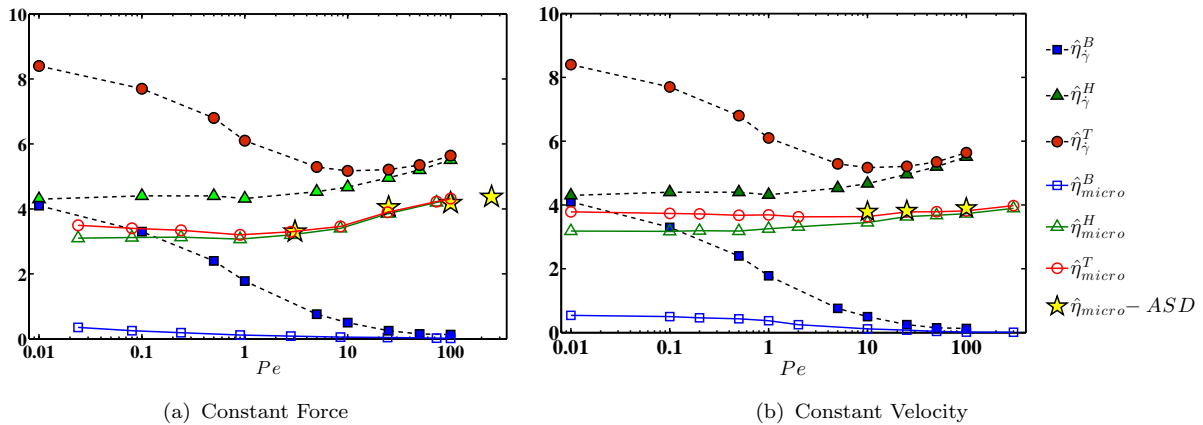

**Figure 15:** *Viscosity as a function of Pe for $\phi = 0.40$ at (a) constant force and (b) constant velocity. The open symbols are the predicted values for microrheology and filled symbols are the results of ASD simulations in shear flow at the equivalent $Pe_{\dot{\gamma}}$. The stars present the total microviscosity from ASD simulation results for both CF and CV.*

## 8 Summary

We have used the discrete-particle simulation method Accelerated Stokesian Dynamics (ASD) to evaluate the microstructure (pair distribution function, $g(\mathbf{r})$) and rheology (viscosity) of Brownian colloidal suspensions in active microrheology (MR). In this technique, a probe particle is pulled through the suspension bath. The work has considered a wide range of flow conditions in terms of the Péclet number $Pe$, and different volume fractions. This is, to our knowledge, the first simulation work that considers detailed hydrodynamic interactions in active MR. We have considered pulling a probe particle with fixed external force (CF) and an approximation to fixed velocity (CV) as analogs to optical tweezer and magnetic bead experiments, respectively. Through simulation we demonstrate that the major difference between these two conditions at $Pe \gg 1$ results from the dispersive motion of the probe under CF conditions; this dispersion is a result of velocity fluctuations associated with the heterogeneity of the environment encountered by the probe as well as thermal fluctuations, with the former dominant at elevated Péclet number for the concentrations studied here. By contrast, moving the probe with a fixed velocity, the CV condition, results in a straight-line trajectory. As a consequence, the pair distribution function in CF demonstrates one major difference compared to CV: when pulling with constant velocity, the length of the zone depleted of bath particles, or wake, formed behind the probe at $Pe \gg 1$ continuously increases with $Pe_U$ to the point that it spanned the entire simulation box; see figure 10. In contrast, the ASD simulations for CF conditions show saturation of the length of the wake (and other far-field structure) with $Pe_f$. We demonstrate that the force-induced dispersion is a direct result of many-body hydrodynamic interactions (HI). This effect has not been considered in previous theories of active microrheology and simulations since they either ignore HI or assume $\phi \ll 1$ and limit interactions to two-particle level.

We corroborate the ASD simulation results with a microscopic theory based on the extension of our theoretical framework for the microstructure and rheology of sheared suspensions [Nazockdast & Morris, 2012b, 2013]. The theory computes $g(\mathbf{r})$ as a solution to the pair Smoluchowski equation and is applicable to suspensions with $\phi \leq 0.40$ over the entire range of $Pe$. The predicted microstructure $g(\mathbf{r})$ has been used here to compute the average mobility of the probe and from this the apparent microrheologically-determined viscosity as a function of $\phi$ and $Pe$. A novel aspect of the theory involves modeling the effect of force-induced dispersion noted just above to be a factor in CF conditions. For this, we followed the simple practice adopted in our work on sheared suspensions [Nazockdast & Morris, 2012b] and modeled the force-induced relative dispersion flux as a diffusion in the pair Smoluchowski equation.

We were not able to obtain statistically meaningful measure of structural distortion in near equilibrium limit, $Pe \ll 1$ from ASD simulations. Thus we have only used our theory to study structure and viscosity in this limit. We find that the distorted microstructures are quite similar for both fixed force and velocity conditions. This is a consequence of the many-body interactions which result in equilibrium structure and a potential of mean force, and thus is not in agreement with the predictions of dilute theories, which predict a distortion twice as large for constant velocity compared to constant force. We demonstrate that including many-body interparticle interactions results in a significantly sharper decay of structural distortion away from the probe, compared to the predictions of the dilute theory at volume fractions as low as $\phi = 0.20$. As a result of this, the structure and viscosity are primarily set by near-contact dynamics of the probe with the bath particles. We show that these interactions are very similar in both CF and CV modes of moving the probe, resulting in qualitatively similar structure and viscosity. We also find that the contribution of Brownian forces to the total viscosity in active microrheology is markedly smaller, by almost an order of magnitude, than in shear flow.

The theoretical predictions of $g(\mathbf{r})$ are in good agreement with ASD results, both near-contact and at larger distances, for the range of $Pe$ and $\phi$ studied here. Most notably, as result of including the effect of force-induced diffusion, the theory can successfully predict the saturation of the wake other structural features away from contact at $Pe_f \gg 1$. Our simulation results and predictions show that near-contact pair microstructure scales similarly with $Pe$ and $\phi$ for CF and CV. We show that the structure at $Pe \gg 1$ is primarily set by the force consumed to pull the particle. This observation is explained by noting that the force-induced diffusivity becomes negligible for near contact configurations; thus the ratio of convection to diffusion fluxes scales similarly for both CF and CV pullings as $Pe^{-1}$ with no strong dependency on $\phi$. As a result the values for viscosity at $Pe \gg 1$ are quantitatively close for CF and CV in predictions and ASD results. In both simulation and theory we find a mild increase of viscosity with the external force/velocity. The extent of this *force-thickening* behavior is, however, smaller than shear thickening in shear flow simulated by Stokesian Dynamics.

We have limited our study here to the case of equal size of probe and bath particles. Previous studies have shown a strong influence of probe to bath particle size ratio on microstructure and microrheology [Almog & Brenner, 1997]. This theory can be extended to consider different size ratios through modification of the pair resistance and mobility functions. In this study of active microrheology only hard-sphere suspensions were considered whereas the theory can be easily extended to colloidal suspensions with other types of interactions.

# References

Almog, Y. & Brenner, H. 1997 Non-continuum anomalies in the apparent viscosity experienced by a test sphere moving through an otherwise quiescent suspension. *Phys. Fluids* **9**, 16–22.

Banchio, A. J. & Brady, J. F. 2003 Accelerated Stokesian Dynamics: Brownian motion. *J. Chem. Phys.* **118**, 10323–10332.

Brady, J. F. & Morris, J. F. 1997 Microstructure of strongly sheared suspensions and its impact on rheology and diffusion. *J. Fluid Mech.* **348**, 103–139.

Carpen, I. C. & Brady, J. F. 2005 Microrheology of colloidal dispersions by Brownian dynamics simulations. *J. Rheol.* **49**, 1483–1502.

Choi, S. Q., Steltenkamp, S., Zasadzinski, J. A. & Squires, T. M. 2011 Active microrheology and simultaneous visualization of sheared phospholipid monolayers. *Nat. Commun.* **2**, 312.

Dasgupta, B. R. & Weitz, D. A. A. 2005 Microrheology of cross-linked polyacrylamide networks. *Phys. Rev. E* **71**, 021504.

Depuit, R. J. & Squires, T. M. 2012*a* Micro-macro-discrepancies in nonlinear microrheology : I . Quantifying mechanisms in a suspension of Brownian ellipsoids. *J. Phys.: Condense. Matter* **24**, 464106.

Depuit, R. J. & Squires, T. M. 2012*b* Micro-macro discrepancies in nonlinear microrheology : II . Effect of probe shape. *J. Phys.: Condense. Matter* **24**, 464107.

Dhont, J. K. G. 1996 *An Introduction to Dynamics of Colloids*. Elsevier.

Gazuz, I., Puertas, A. M., Voigtmann, T. & Fuchs, M. 2009 Active and nonlinear microrheology in dense colloidal suspensions. *Phys. Rev. Lett.* **102**, 248302.

Gnann, M. V., Gazuz, I., Puertas, A. M., Fuchs, M. & Voigtmann, T. 2011 Schematic models for active nonlinear microrheology. *Soft Matter* **7**, 1390–1396.

Habdaas, P., Schaar, D., Levitt, A. C. & Weeks, E. A. 2004 Forced motion of a probe particle near the colloidal glass forced motion of a probe particle near the colloidal glass transition. *Europhys. Lett.* **67**, 477–483.

Khair, A. & Brady, J. F. 2006 Single particle motion in colloidal dispersions: a simple model for active and nonlinear microrheology. *J. Fluid Mech.* **557**, 73–117.

Khair, A. & Brady, J. F. 2008 Microrheology of colloidal dispersions: Shape matters. *J. Rheol.* **52**, 165–196.

Leshansky, A. M. & Brady, J. F. 2005 Dynamic structure factor study of diffusion in strongly sheared suspensions. *J. Fluid Mech.* **527**, 141–169.

Leshansky, A. M., Morris, J. F. & Brady, J. F. 2008 Collective diffusion in sheared colloidal suspensions. *J. Fluid Mech.* **597**, 305–341.

Lionberger, R. A. & Russel, W. B. 1997 A Smoluchowski theory with simple approximations for hydrodynamic interactions in concentrated dispersions. *J. Rheol.* **41**, 399–425.

MacKintosh, F. C. & Schmidt, C. F. 1999 Microrheology. *Curr. Opin. Colloid Interface Sci.* **4**, 300–307.

Mason, T. G. & Weitz, D. A. 1995 Optical measurements of frequency-dependent linear viscoelastic moduli of complex fluids. *Phys. Rev. Lett.* **74**, 1250–1253.

Mohan, L., Cloitre, M. & Bonnecaze, R. T. 2014 Active microrheology of soft particle glasses. *J. Rheol.* **58**, 1465.

Nazockdast, E. & Morris, J. F. 2012*a* Effect of repulsive interactions on structure and rheology of sheared colloidal dispersions. *Soft Matter* **8**, 4223–4234.

Nazockdast, E. & Morris, J. F. 2012*b* Microstructural theory and the rheology of concentrated colloidal suspensions. *J. Fluid Mech.* **713**, 420–452.

Nazockdast, E. & Morris, J. F. 2013 Pair-particle dynamics and microstructure in sheared colloidal suspensions: simulation and smoluchowski theory. *Phys. Fluids* **25**, 070601.

Pearson, D. S. & Shikata, T. 1994 Viscoelastic behavior of concentrated spherical suspensions. *J. Rheol.* **38**, 601–616.

Rice, A. S. & Lekner, J. 1965 On the equation of state of the rigid-sphere fluid. *J. Chem. Phys.* **42**, 3559–3565.

Sierou, A. & Brady, J. F. 2001 Accelerated Stokesian Dynamics simulations. *J. Fluid Mech.* **448**, 115–146.

Sierou, A. & Brady, J. F. 2002 Rheology and microstructure in concentrated noncolloidal suspensions. *J. Rheol.* **46**, 1031–1056.

Squires, T. M. 2008 Nonlinear microrheology : Bulk stresses versus direct interactions. *Langmuir* **24**, 1147–1159.

Squires, T. M. & Brady, J. F. 2005 A simple paradigm for active and nonlinear microrheology. *Phys. Fluids* **17**, 073101.

Squires, T. M. & Mason, T. G. 2010 Fluid mechanics of microrheology. *Ann. Rev. Fluid Mech.* **42**, 413–438.

Sriram, I., Meyer, A. & Furst, E. M. 2010 Active microrheology of a colloidal suspension in the direct collision limit. *Phys. Fluids* **22**, 062003.

Swan, J. W. & Zia, R. N. 2013 Active microrheology: Fixed-velocity versus fixed-force. *Phys. Fluids* **25**, 083303.

Swan, J. W., Zia, R. N. & Brady, J. F. 2014 Large amplitude oscillatory microrheology. *J. Rheol.* **58** (1), 1 – 41.

Valentine, M. T., Kaplan, P. D., Thota, D., Crocker, J. C., Gisler, T., Prud, R. K., Beck, M. & Weitz, D. A. 2001 Investigating the microenvironments of inhomogeneous soft materials with multiple particle tracking. *Phys. Rev. E* **64**, 61506.

Verma, R., Crocker, J. C., Lubensky, T. C. & Yodh, A. G. 1998 Entropic colloidal interactions in concentrated DNA solutions. *Phys. Rev. Lett.* **81** (18), 4004–4007.

Voigtmann, T. & Fuchs, M. 2013 Force-driven micro-rheology. *Euro. phys. J.* **232**, 2819–2833.

Waigh, T. A. 2005 Microrheology of complex fluids. *Rep. Prog. Phys.* **68**, 685–742.

Willenbacher, N. & Oelschlaeger, C. 2007 Dynamics and structure of complex fluids from high frequency mechanical and optical rheometry. *Curr. Opin. Colloid. Interface Sci.* **12**, 43–49.

Wilson, L. G., Harrison, A. W., Schofield, A. B., Arlt, J. & Poon, W. C. K. 2009 Passive and Active Microrheology of Hard-sphere Colloids. *J. Phys. Chem. B* **113**, 3806–3812.

Wirtz, D. 2009 Particle-tracking microrheology of living cells: Principles and applications. *Ann. Rev. Biophys.* **38**, 301–326.

Zia, R. N. & Brady, J. F. 2012 Microviscosity, microdiffusivity, and normal stresses in colloidal dispersions. *J. Rheol.* **56**, 1175–1208.